# Current water trapping micro-habitats on the surface of Mars

Anna Bognar[1,2], Bernadett D. Pal[1,2], Akos Kereszturi[2,3]

[1] ELTE Eötvös Loránd University, Institute of Physics and Astronomy, Department of Astronomy, 1117 Budapest, Pázmány Péter sétány 1/A, Hungary

[2] HUN-REN Research Centre for Astronomy and Earth Sciences, Konkoly Observatory,

MTA Centre of Excellence, Konkoly Thege Miklós út 15-17., H-1121 Budapest, Hungary

[3] European Astrobiology Institute, Strasbourg, France.

**Highlights:**

The study presents a conceptual approach to potentially habitable micro-environments on the present-day Martian surface.

Salt-crystal cracks could open at night due to low-temperature-induced contraction.

Microscopic liquid by deliquescence could seep along these cracks to crystal interiors.

Daytime warming may contribute to crack closure and temporary liquid retention within the crystal interior.

The concept is evaluated using calculations for NaCl and $Ca(ClO_4)_2$, and the process may be feasible under the examined conditions, although water-retention and diffusion effects require further investigation.

**Abstract**

This paper presents a conceptual model exploring a possible mechanism by which liquid water could potentially be maintained in direct contact with hypothetical microorganisms during warm daytime periods on Mars. Hygroscopic salts detected on the surface are capable of binding water vapor from the atmosphere and creating a liquid solution, however during daytime they could dry out again if they have direct contact with the dry atmosphere. In this work a model is presented, where crystalline fractures close after the cold and wet night, capturing the nighttime condensed liquid inside hygroscopic salt crystals for the daytime hours. In the evening and morning hours, the relative humidity can reach a level where, in the case of certain salts (e.g., $Ca(ClO_4)_2$), a thin surface liquid layer can form on mineral surfaces. Due to daily temperature fluctuations, the salts undergo significant thermal changes. These mechanical stresses can lead to crack formation, which can be closed by daytime growth and then opened again by contraction during cooler nights. However, crack closure does not necessarily imply complete sealing, and the long-term retention of liquid water likely depends on additional thermo-mechanical factors and vapor transport conditions. Previous numerical studies suggest that thermally induced stresses under Martian conditions may indeed be sufficient to support crack formation and subcritical crack growth; however, the detailed evolution of crack aperture likely depends on additional factors including mineral anisotropy, pre-existing microfractures, and local thermo-mechanical conditions. Along these cracks a deliquescence-produced microscopic liquid layer could penetrate the interior of salt crystals. Hygroscopic salts relevant to the Martian environment—including NaCl, $CaCl_2$, $Ca(ClO_4)_2$, gypsum, and $MgSO_4$ polymorphs—exhibit measurable thermally induced volume changes under diurnal temperature variations. Estimated relative volume changes span from ~0.5–0.9% for crystalline phases (e.g., $MgSO_4$, gypsum) up to ~1.6% for NaCl and ~3.7–6.9% for $CaCl_2$ solutions, suggesting that these effects may be significant in shaping near-surface microenvironments. However, direct thermal expansion data for hydrated calcium perchlorates under Mars-relevant conditions remain limited, and therefore any analogous behaviour inferred for $Ca(ClO_4)_2$ should presently be regarded as preliminary and hypothetical. If a broadly similar expansion–contraction behaviour occurs in more hygroscopic salts such as $Ca(ClO_4)_2$, such materials could represent possible transient microenvironments on Mars today. If hypothetical organisms are inside such crystals, because of the daily expansion-contraction cycle, liquid water could be maintained there in the elevated temperature daytime period. Evaluating the spatial distribution of deliquescence and observed salt occurrence on Mars, nighttime microscopic liquid can occur for a cumulative total of approximately 100-130 sols in a Martian year in the vicinity of Acidalia Planitia at several salt occurrence locations. At a depth of approximately 2-3 mm in the soil, microorganisms may receive partial UV shielding, while potentially still receiving limited visible light, depending on the optical properties of the surrounding material. In summary, the crystals could potentially provide transient liquid water to a hypothetical bacterium through an internal crack even during daytime dryness, making these locations a possible candidate environment for photosynthetic organisms on Mars today. Therefore, the present study should be regarded primarily as a conceptual feasibility analysis and hypothesis for transient near-surface brine retention within hygroscopic salt crystals under present-day Martian conditions.

# 1. Introduction

Present-day Mars poses a major challenge for habitability because periods with elevated relative humidity usually coincide with very low temperatures, while warmer daytime conditions are generally extremely dry. In this work we discuss only a specific aspect of this topic, exploring in greater detail the current habitability of Mars inside near surface salt crystals, providing elevated humidity at daytime together with elevated temperature. While daytime temperature on Mars may become more favourable for metabolic activity in organisms analogous to certain terrestrial extremophiles, the extremely dry atmosphere and limited water availability would strongly inhibit biological activity. In contrast, nighttime conditions may permit atmospheric saturation and transient condensation of $H_2O$ (Titov, 2002), including microscopic liquid formation (Zorzano et al., 2009; Martín-Torres et al., 2015), although temperatures during these periods are generally too low for active metabolism. The research question behind this work is how nighttime condensed water could stay inside partially closed near-surface fractures during the elevated daytime temperature despite the dryness, potentially supporting temporary liquid retention and possible metabolism.

In this work we present a model that explores whether humidity and water acquired at night could be retained into the warmer daytime period. Below the background of this model is presented, considering the current conditions and materials on the surface of Mars. It should be noted that this model does not consider ionizing particle radiation, which represents a major limitation for potential near-surface habitability on present-day Mars. We outline how nighttime condensation could provide microscopic liquid to the interior of certain salts (calculated model results), and later how these voids may become partially closed during warmer daytime conditions to keep the $H_2O$ inside during the warmer daytime (thermal induced volume change). The proposed model is constrained by the lack of several relevant physical-property data. Therefore, the present study relies partly on theoretical expectations and laboratory measurements obtained for selected Mars-relevant minerals, which should be regarded as preliminary approximations rather than direct representations of real Martian materials.

Deliquescence refers to the phenomenon whereby a salt changes from a solid to a liquid state due to an increase in the relative humidity of the atmosphere. Mineral deliquescence is a process in which certain salts absorb atmospheric water vapor at specific relative humidity levels, forming a liquid solution on the surface of their crystals (Davila et al., 2010). The salt binds the water vapor from its environment, causing a thin liquid phase layer to appear on the surface of the mineral, ultimately resulting in a saturated salt solution. The reverse process is called salt efflorescence, when a solid crystal forms as a result of a decrease in RH (relative humidity). The process is influenced by a number of factors, including ambient temperature, atmospheric humidity, and the hydration state of the salt.

Even if environmental conditions on Mars are periodically favourable, the long-term survival of liquid water remains a fundamental challenge. However, there is a significant amount of **hygroscopic salt** on the surface of Mars, which may promote water uptake through deliquescence under favourable conditions, while the resulting brines may reduce evaporation because of their lower saturation vapor pressure (Martínez and Renno, 2013). Some experiments show that even the small amount of moisture produced during deliquescence may be sufficient to reactivate methanogenic metabolism in certain halotolerant archaea under Mars-like conditions (Maus et al., 2020). An important component of the Martian water cycle is water vapor in the air and its interaction with the surface. The amount of water vapor is influenced by the sublimation and freezing of ice at the poles, its interaction with surface regolith and surface ice layers, and the movement of wind and air currents. The amount of water vapor in the air reaches its maximum in the morning and at noon. Sunlight heats the surface (e.g., regolith), causing water vapor to be released. In the afternoon, the emitted water vapor remains near the surface. Subsequently, as the temperature decreases and the surface cools, the relative humidity increases and the water vapor can bind to the regolith again. In other words, the regolith actively participates in the water cycle (Titov, 2002). It is important to note that although the absolute amount of water vapor reaches its maximum in the late morning and around noon, the relative humidity does not peak at this time. Relative humidity increases as temperature decreases, therefore it reaches its maximum during the late night and early morning hours. In the following section, we overview the occurrence of water-trapping salts and the conditions required for deliquescence and liquid water formation. We further discuss how thermally induced volume changes in such salts could potentially contribute to fracture closure and temporary liquid retention within crystalline microenvironments. The aim of this work is therefore to propose and evaluate a conceptual mechanism by which hygroscopic salt crystals may temporarily stabilize liquid water under present-day Martian conditions.

# 1.1 Important Martian surface conditions

Several hygroscopic salts have been identified on the surface of Mars which are able to trap $H_2O$ from the atmosphere. Osterloo et al. (2008) investigated the occurrence of chloride-containing minerals on Mars based on infrared spectra. The presence of these minerals has already been detected by several instruments, including Mars Odyssey THEMIS, Mars Global Surveyor, and Mars Reconnaissance Orbiter (Osterloo et al., 2008 and Osterloo et al., 2010). It can be observed that such chloride deposits show significant emissivity in the 1260-900 $cm^{-1}$ range, which makes them distinguishable from other surface materials. Individual deposits are relatively small, commonly ranging from about 1 to 25 $km^2$, but they are widespread globally. These deposits can be observed in separate, irregular patches, craters, or even winding valleys probably at outcrops of formerly deposited salts. Based on THEMIS measurements, 200 such deposits have been detected, mainly in low-albedo, mid- and low-latitude areas of the southern highlands (Osterloo et al., 2008).

Observations suggest that most deposits were formed by the evaporation of saline solutions from water collected in craters or depressions, suggesting that surface and subsurface water may have once been present. In situ wet chemistry analyses performed by the Phoenix lander showed that Martian soil solutions contained approximately 10 mM dissolved salts, including 0.4–0.6% perchlorate ($ClO_4^-$) by mass, along with chloride, bicarbonate, and possibly sulphate ions (Hecht et al., 2009). Carter et al. (2013) conducted similar research, identifying more than 990 mineral occurrences. In their case, more than 70% of the sites are located near impact craters, where material ejected by the explosion exposed deeper layers. In addition, they can also be found along slopes and ditches formed by erosion. A significant proportion of the sites identified in this way are located near the equator, specifically within $\pm 40^{o}$ latitude (Carter et al., 2013). Based on various measurements, silicates, sulphates, and chlorides were primarily observed in the areas studied (Osterloo et al., 2008; Osterloo et al., 2010 and Carter et al., 2013). The global distribution of hydrous mineral detections identified by OMEGA and CRISM is shown in Fig. 1, with many occurrences concentrated in equatorial regions. Figure 2 shows the global distribution of chloride-bearing deposits identified by Osterloo et al. (2008), providing a complementary view of salt occurrences on Mars.

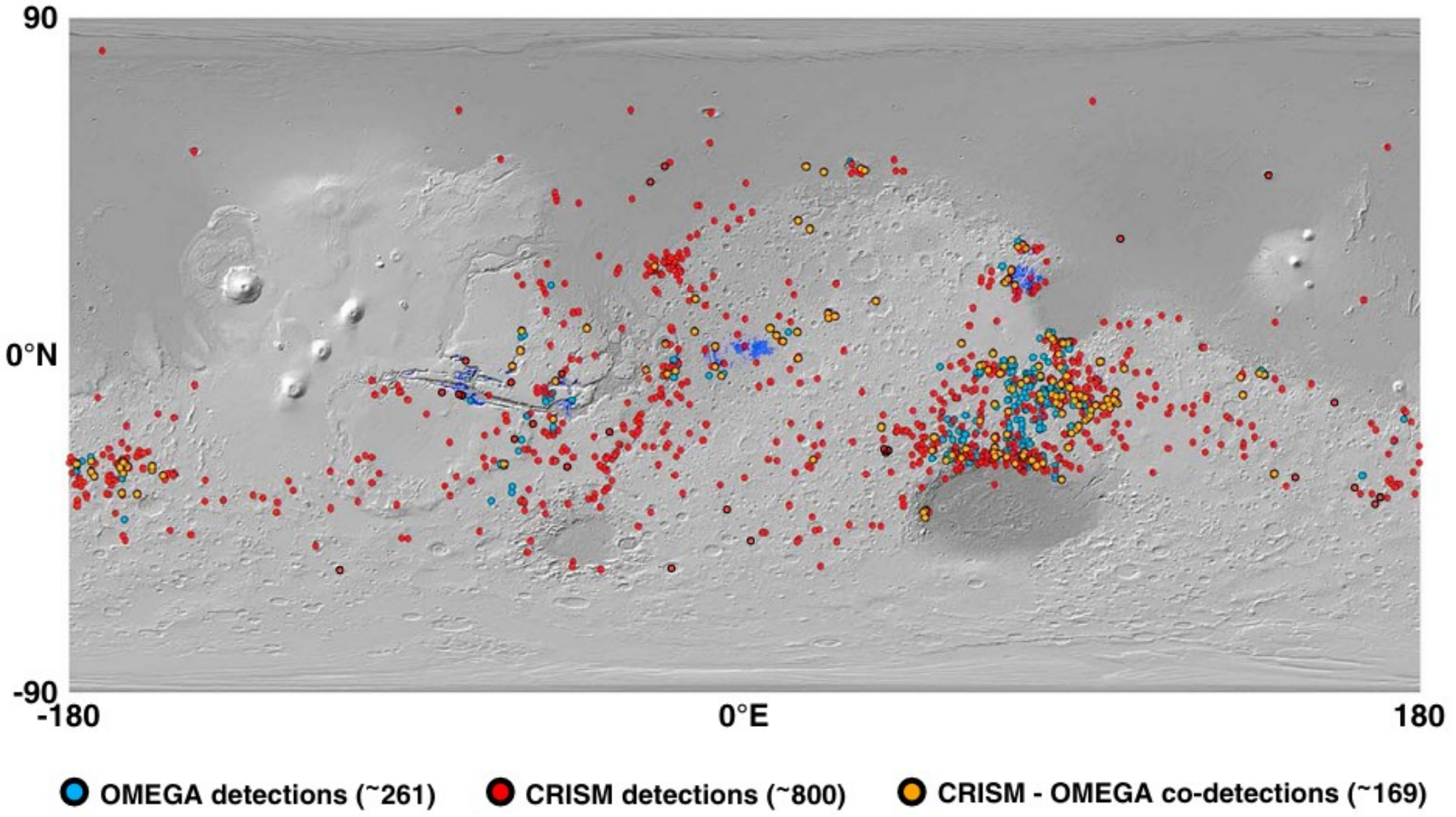

1. *Figure: Global distribution of hydrous mineral detections on Mars. Symbols mark locations where hydrous minerals have been identified from orbital spectroscopy: CRISM detections are shown in red, OMEGA detections in blue, and sites observed by both instruments in orange. For CRISM data, a single exposure is counted per observation regardless of the number of hydrous mineral species detected (Carter et al. 2013).(For interpretation of the references to colour in this figure legend, the reader is referred to the web version of this article.)*

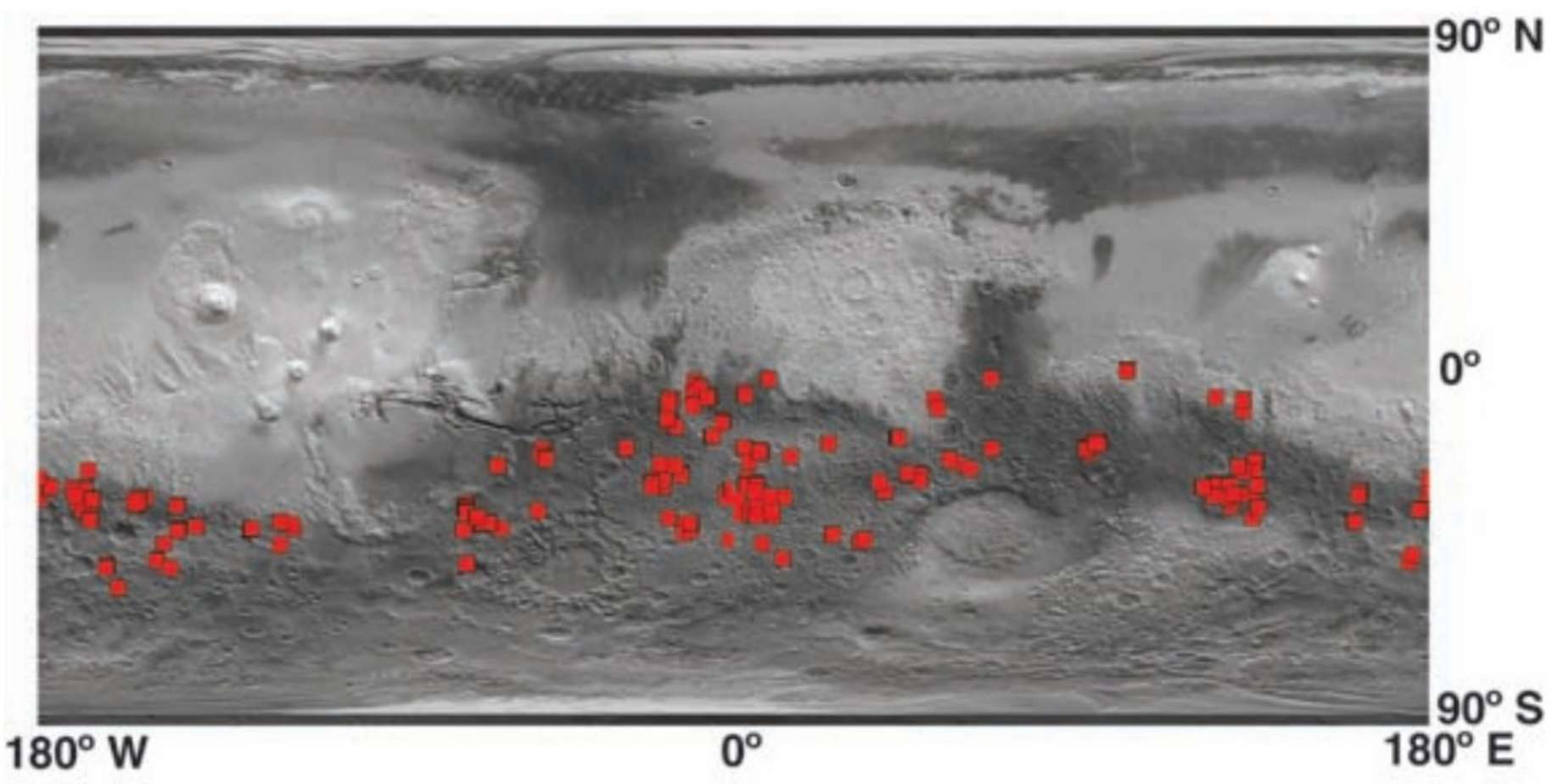

2. *Figure: Global distribution of chloride-bearing deposits on Mars mapped from orbital data (Osterloo et al., 2008). These deposits indicate the past presence of evaporitic environments.*

Among these minerals, hygroscopic salts are of particular interest because of their unique ability to absorb atmospheric water vapor and liquefy as a result (Gough et al., 2016). However, for this process to occur, the ambient temperature and atmospheric humidity must reach the threshold values characteristic of the salt in question. Thanks to this process, liquid water may transiently form locally on the surface of Mars under highly salt-dependent conditions (Martín-Torres et al., 2015; Zorzano et al., 2009; Pál and Kereszturi, 2020). It is therefore important to examine how such a crystal could provide moisture for a hypothetical living organism while protecting it from extreme weather conditions and surface UV radiation.

**Annual temperature and atmospheric humidity fluctuations** lead to changes in many parameters influencing the deliquescence process, and thus the possibility of microscopic liquid formation on mineral surfaces.

## 1.2 Water formation by Martian salt crystals, deliquescence

In this section we summarize specific data and properties of Mars surface materials, which influence the temperature related water budget and humidity behaviour, as well as the volume of various candidate hygroscopic salts. The results gained using these background data are presented in the Results section.

When studying salts and salt mixtures found on Mars, it is important to thoroughly investigate the phenomenon of deliquescence, as it is closely related to the liquid water that can be found on the planet today. One of the special properties of hygroscopic salts is their ability to bind atmospheric water vapor (Gough et al., 2016) below saturation relative humidity. Under certain conditions, this process forms a liquid, microscopic, saturated salt solution on their surface, which can create liquid water on the surface of Mars on a small scale despite the general dryness (Martín-Torres et al., 2015, Zorzano et al., 2009) especially at night, when relative humidity is elevated.. The deliquescence of hygroscopic minerals such as chloride salts may therefore provide a local and transient source of liquid water under present-day Martian conditions (Davila et al., 2010). One example of such hygroscopic salts is calcium perchlorate, which is highly deliquescent and may exist on the Martian surface under present-day environmental conditions (Nuding et al., 2014).

One of the most important influencing factors is the deliquescence relative humidity (DRH), which indicates the minimum relative humidity required for the salt to begin to transform from a solid state to a liquid phase. The relative humidity at which this process begins is known as the deliquescence relative humidity (DRH), and each mineral has its own characteristic DRH value (Davila et al., 2010). The DRH value is different for each salt and generally increases with rising temperature. Laboratory measurements show that the deliquescence relative humidity of calcium perchlorate can vary widely, ranging from approximately 5% to 55% depending on temperature and hydration state (Nuding et al., 2014). Experimental results indicate that different hydration phases of $Ca(ClO_4)_2$, including anhydrous and hydrated forms, exhibit different DRH values, with more highly hydrated phases generally deliquescing at higher relative humidity (Nuding et al., 2014). The DRH values of perchlorates and chlorides on Mars can vary significantly, and these values also change when the salts form different mixtures with each other. In the case of mixtures, the percentage ratio of each component also plays an important role in the development of the process (Gough et al., 2014). When the environmental parameters are right for deliquescence, a very thin, liquid layer first forms on the surface of the mineral salts. The amount of solid salt gradually decreases during the process, and eventually a saturated salt solution is formed (Martín-Torres et al., 2015). This solution can exist on the surface of Mars for several hours during the morning and evening periods. The conditions for its formation are influenced by atmospheric conditions, seasonal changes, and the geographical location of the area under investigation (Pál and Kereszturi, 2017).

The basic condition for deliquescence is that the relative humidity (RH) of the atmosphere reaches or exceeds the DRH value characteristic of the given salt. However, it is important to note that as the temperature increases, the DRH value of the salt also increases, meaning that at higher temperatures, a higher relative humidity is required for deliquescence to occur (Wang et al., 2011). In the case of mixtures, the deliquescence behaviour may differ substantially from that of the individual components because mixed systems can exhibit eutonic behaviour and non-ideal thermodynamic interactions (Gough et al., 2014). As a result, the effective DRH of the mixture may be lower than the DRH values of the pure salts.

To understand deliquescence accurately, it is important to consider ion–water interactions and hydration processes at the salt surface. Experimental studies have shown that water molecules can accumulate on hygroscopic salt surfaces and form thin hydrated surface layers even below the thermodynamically predicted deliquescence threshold (Fauré et al., 2023). The properties and stability of these layers depend on ion hydration and solvation processes. In addition, adsorbed water may remain trapped within microscopic pore spaces of the crystal structure under low-relative-humidity conditions (Fauré et al., 2023). In the case of hygroscopic salt crystals, water uptake may initially form a thin hydrated surface layer on the mineral surface, which under favourable conditions can further develop into a partially liquid deliquescent brine layer.

# 1.3 Diurnal Temperature Effects

An important but poorly evaluated parameter is the volume change. In a 2020 study, Angell et al. (2020) used NASA's Mars Odyssey spacecraft to examine daily and seasonal changes in surface temperature in Gale Crater, located near the equator. The probe is equipped with a THEMIS instrument, which can measure the surface heat emission in the infrared range, from which the surface temperature can be determined. The temperature measured between 4 and 6 a.m. local time ranged from 185 K to 205 K, while measurements taken between 4 and 6 p.m. ranged from 215 K to 240 K. Based on these measurements, it can be concluded that the daily temperature variation in Gale Crater is at least 55 K (Angell et al., 2020). These Gale Crater measurements are presented here primarily as an illustrative example of the magnitude of daily Martian surface temperature fluctuations based on well-constrained in situ observations, and were not used directly in the later calculations.

However, it is important to examine in more detail how much change the daily cycle causes when comparing the hottest and coldest hours. This parameter was examined by Kuziakina et al. (2019) and Kuti and Kereszturi (2007) . Kuti and Kereszturi (2007) conducted measurements around the northern summer solstice. Since the salt crystals of interest to us were found in the equatorial region, approximately between 40 degrees latitude and the equator, both north and south, it is worth examining the values there. In all of the regions studied, these areas had the most significant daily temperature changes. The average daytime maximum temperature was 273 K, while the nighttime minimum was around 150 K in the vicinity of the equator (Kuti and Kereszturi, 2007). Similar temperature values were measured by Kuziakina et al. (2019) in Schiaparelli Crater and the Elysium Planum region, both located near the equator. In Schiaparelli Crater, the average daytime temperature was 275 K, while the nighttime temperature was 163 K. In the Elysium Planum, the daytime temperature was slightly higher, at 279 K, while the nighttime temperature reached 193 K (Kuziakina et al., 2019). The two studies obtained similar values for daily temperature fluctuations. Due to this change, the volume of the crystal increases or decreases, which can cause mechanical changes (Drebushchak, 2020 and Wallace, 1972).

## 1.4 Temperature and Humidity Variability on Mars

Due to the extreme weather conditions on Mars, significant changes in temperature and humidity can be observed during a daily cycle.
The amount of atmospheric water vapor on Mars is primarily controlled by the condensation and sublimation of the polar caps, exchange with the regolith and surface frost, and advective transport by the general circulation (Titov, 2002).
Based on measurements from the Viking MAWD (Mars Atmospheric Water Detector) experiment, the atmospheric water vapor column density typically varies between 10 and 20 precipitable microns (pr. µm) on an annual scale. During northern summer, this value increases significantly due to the sublimation of the residual northern polar cap, whereas the increase is less pronounced in the southern hemisphere (Titov, 2002). Diurnal variations in water vapor are also significant. The maximum column abundance is generally observed around noon, when solar heating of the surface enhances the release of water vapor from the regolith. As surface temperature decreases, the relative humidity increases and water is re-adsorbed into the near-surface layers. The magnitude of diurnal variation is typically on the order of ~10 pr. µm, although it can vary depending on location (Titov, 2002). These diurnal humidity variations are especially important for near-surface hygroscopic salts, because the nighttime increase in relative humidity may create temporarily favourable conditions for local deliquescence processes. These pronounced diurnal and seasonal variations directly influence the exchange of water between the regolith and the atmosphere. Estimates suggest that up to ~10 cm of the Martian soil participates in the annual exchange, while the uppermost ~1 cm of the regolith is actively involved in the diurnal exchange cycle. Despite its small thickness, this near-surface layer can store as much as 100–1000 pr. µm of water (Titov, 2002). The proposed salt-crystal microhabitats are assumed to occur within this near-surface exchange layer, where periodic interaction between the atmosphere and the regolith may provide temporary access to atmospheric water vapor.

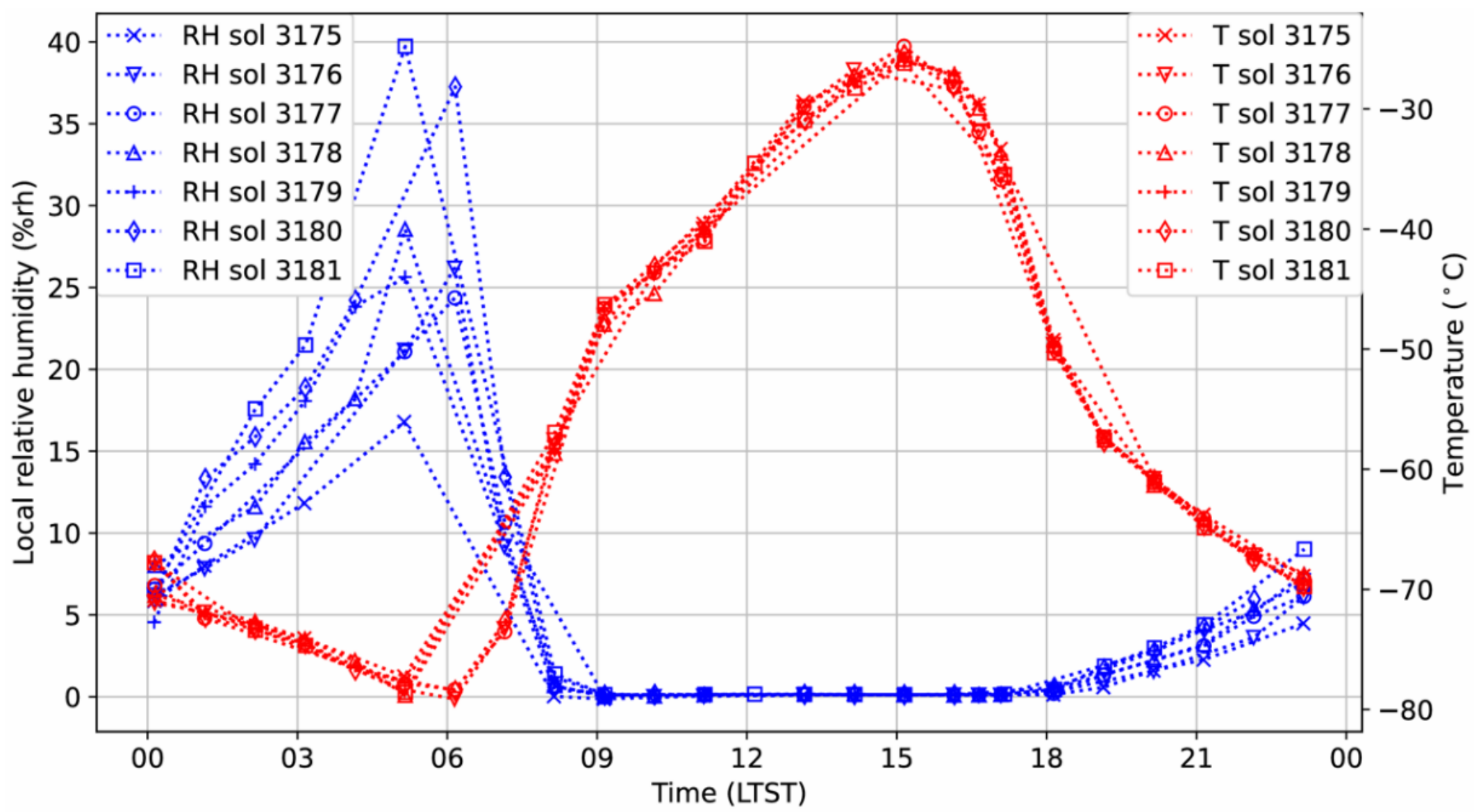

3. Figure: Diurnal variations in near-surface relative humidity (RH) and air temperature over seven consecutive Martian sols (3175–3181) in Gale Crater, as a function of Local True Solar Time (LTST). Relative humidity peaks in the early morning (~17–40%) at minimum temperatures and drops to near zero during daytime, demonstrating a strong inverse relationship between temperature and RH (Hieta et al., 2025). Adapted from Hieta et al. (2025).

Figure 3, adapted from Hieta et al. (2025), shows the daily variations in local relative humidity (RH) and air temperature over seven consecutive Martian days (sol 3175–3181). The horizontal axis of the figure shows Local True Solar Time (LTST) from 0 to 24 h, while the vertical axis on the left represents relative humidity and the axis on the right represents temperature. The blue

markers and curves show relative humidity values, while the red markers represent temperature data; each curve corresponds to a separate Martian day (Hieta et al., 2025). The data show a well-defined daily cycle. Relative humidity gradually increases during the night and early morning hours, typically reaching a maximum of 17–40% in the period before sunrise. This period coincides with daily minimum temperatures, which range from approximately −75 to −80 °C. After sunrise, relative humidity drops sharply and falls to practically zero within a few hours. During the daytime, between approximately 09:00 and 17:00 LTST, the RH value remains at ~0%, while the temperature reaches its daily maximum. In the late afternoon and evening hours, as the temperature drops, relative humidity begins to rise again, and higher RH values re-emerge during the night (Hieta et al., 2025). Figure 3 clearly illustrates the strong inverse relationship between temperature and relative humidity under Martian conditions rather than simultaneous deliquescence conditions for all salts discussed above. Under the RH conditions shown here, deliquescence would likely be restricted to the most hygroscopic salts, particularly certain perchlorate phases. The figure clearly demonstrates that the lowest temperatures correspond to the highest RH values, while daytime warming is accompanied by a drastic decrease in relative humidity (Hieta et al., 2025).

# 2. Methods

In this section first the physical background is presented on how the volume changes by temperature modification, and the calculation of temperature change using model based approach secondly. We performed volume change calculations for various salts between 150 K and 300 K. During the calculations, we applied the approximation commonly used for linear thermal expansion, which allowed us to determine the percentage change in volume. We then used modelled Martian climate data to examine the possibility of deliquescence for pure hygroscopic salts, and calculated the cumulative sols in a year when brines could theoretically emerge.

## 2.1. Volume change as a function of temperature

The volume of materials changes as a result of temperature changes; this is called thermal expansion. The primary cause of thermal expansion is a change in the vibrations of the atoms in the crystal lattice. As the temperature rises, the amplitude of the atoms' vibrations increases, leading to a change in the parameters of the lattice. This process results in an increase in the volume of the crystal, which is called volumetric thermal expansion. Below, a simple numerical estimate of the consequences of this is presented. First, let us consider the change in volume of a regular cube with side length *L* as a function of temperature change Δ*T.* This can be traced back to a combination of linear thermal expansion:

$$\Delta L = L_0 \cdot \alpha \cdot \Delta T$$

In terms of volume, $V_0 = L_0^3$, from which we can obtain the following relationship for the change in volume of a regular shape with sides of length *L*:

$$V = V_0 \cdot (1 + \alpha \cdot \Delta T)^3 \cong V_0 \cdot (1 + 3\alpha \cdot \Delta T) = V_0 \cdot (1 + \beta \cdot \Delta T)$$

where $\alpha$ is the linear thermal expansion coefficient and $\beta$ is the volumetric thermal expansion coefficient, see (Drebushchak, 2020). In this case, the quadratic and cubic terms are negligible, since the value of $\alpha$ is small and, as a result, these terms are orders of magnitude smaller. $\beta$ can exhibit different behaviors as a function of temperature, it can increase or decrease, its change can be positive or negative, and it can occur without a breakpoint or with a sharp change at certain temperatures.

## 2.2. The climate model

To determine the possibility of deliquescence at a given location and time, we used data acquired from the research team of the LMDZ GCM (Laboratoire de Météorologie Dynamique Mars General Circulation Model) that modelled the 29th Martian year (Forget et al. 1999, Navarro et al. 2014). The modeling grid is 64 × 48 × 49 in longitude, latitude, altitude, respectively, corresponding to grid boxes in the order of 330 × 220 km near the equator. The first altitude layer of the model is approximately 4 m above surface level, as well as there are separate surface variables (for example, surface pressure, surface temperature). The 29th Martian year does not include a global dust storm. The modelled data was created in the “Climatology” dust scenario setting of the model, which is built by averaging dust scenarios between the 24th and 31st Martian years (excluding the data of global dust storms during the 25th and 28th years). The model output is for 2 h intervals, and the output data was processed with our own $C^{++}$ codes (calculate RH, filter for deliquescence conditions, filter and restructure data to create maps with same local times on all longitudes), then was converted to NetCDF files and visualized with the NASA Panoply software. We had data modelled for the entire 29th Martian year and created filtered NetCDF databases to check the deliquescence probability at 2 h intervals between 5 PM and 5 AM local time.

The relative humidity with respect to ice ($Q_{sat\ i}$) and with respect to liquid ($Q_{sat\ l}$) was calculated from the modelled variables of saturation water vapor volume mixing ratio, surface temperature, and surface pressure, with an equation based on the Goff-Gratch equation (Goff and Gratch, 1946; List 1951).

$$Q_{sat\ i} = 100/P \cdot 10^{2.07023 - 0.00320991\ T - (2484.896/T) + 3.56654 \log (T)}$$

$$Q_{sat\ l} = 100/P \cdot 10^{23.8319 - (2948.964\ /\ T) - 5.028 \log (T) - 29810.16 \exp (-0.0699382\ T) + 25.21935} \cdot \exp (-2999.924\ /\ T)$$

$RH_{i,l} = Q_0 / Q_{sat\ i,l}$ where P is the surface pressure (Pa), T is the surface temperature (K), and $Q_0$ is the water vapor volume mixing ratio (mol/mol), all of which are direct outputs of the LMDZ GCM model.

## 2.3 Deliquescence probability

Ideal conditions were identified based on the eutectic temperature and water activity of the pure hygroscopic salts $Ca(ClO_4)_2$ ($T_{eut}$ = 199 K, DRH = 0.51 from Pestova et al. 2005) and $CaCl_2$ ($T_{eut}$ = 223 K, DRH = 0.62 from Gough et al. 2016), We focused on late night and early morning hours, when the relative humidity (RH) increases and deliquescence is more likely (Pál et al., 2017). A location and time were flagged as ideal when RH with respect to liquid was equal to or higher than the salt-specific water activity, while RH with respect to ice remained below 145 % in the case of all salts examined. This is because if RH with respect to ice is much higher than 145%, most likely nucleation will occur instead of liquefaction (Rivera-Valentin et al. 2018).

# 3. Results

In this section the temperature based volume change for Mars is presented first, then the proposed process is presented how hypothetical microorganisms could get wet and later in daytime keep the humidity inside the crystal.

Changes in crystal volume are temperature-dependent, which can cause mechanical effects, especially during freezing and melting cycles (Becker et al., 1905). As a result of the nighttime temperature drop, the salt crystals begin to shrink. The change in volume (or other processes) can cause cracks inside them, as tension is generated in the material (Skarbek et al., 2018 and Fletcher et al., 2001) and the cracks may open up further during shrinkage. Cracks may develop in rocks and other solid materials when the applied stress exceeds a given critical threshold. This threshold is determined, among other factors, by the tensile strength of the material and the Griffith critical energy release rate. Rapid temperature variations associated with diurnal thermal cycles may generate thermal stresses within the regolith and rock materials. (Eppes et al., 2015).

Based on numerical models, under Martian conditions, surface temperature variations induced by solar irradiation may be sufficient to generate thermal stresses that could contribute to crack formation and subcritical crack growth. However, this process is highly complex and is influenced by several factors, including the thermal anisotropy of minerals, the optical properties of mineral grains, crystal habit, pre-existing flaws, grain size, elastic properties, thermal gradients, hydration/dehydration state, confinement by surrounding regolith, and fracture roughness and the presence of pre-existing microcracks (Eppes et al., 2015). A quantitative treatment of these parameters remains beyond the scope of the present conceptual model, however some rough estimations have been included.

The temperature dependence of volumetric thermal expansion for several crystalline materials is illustrated in Fig. 4, which shows how expansion coefficients vary across a wide temperature range.

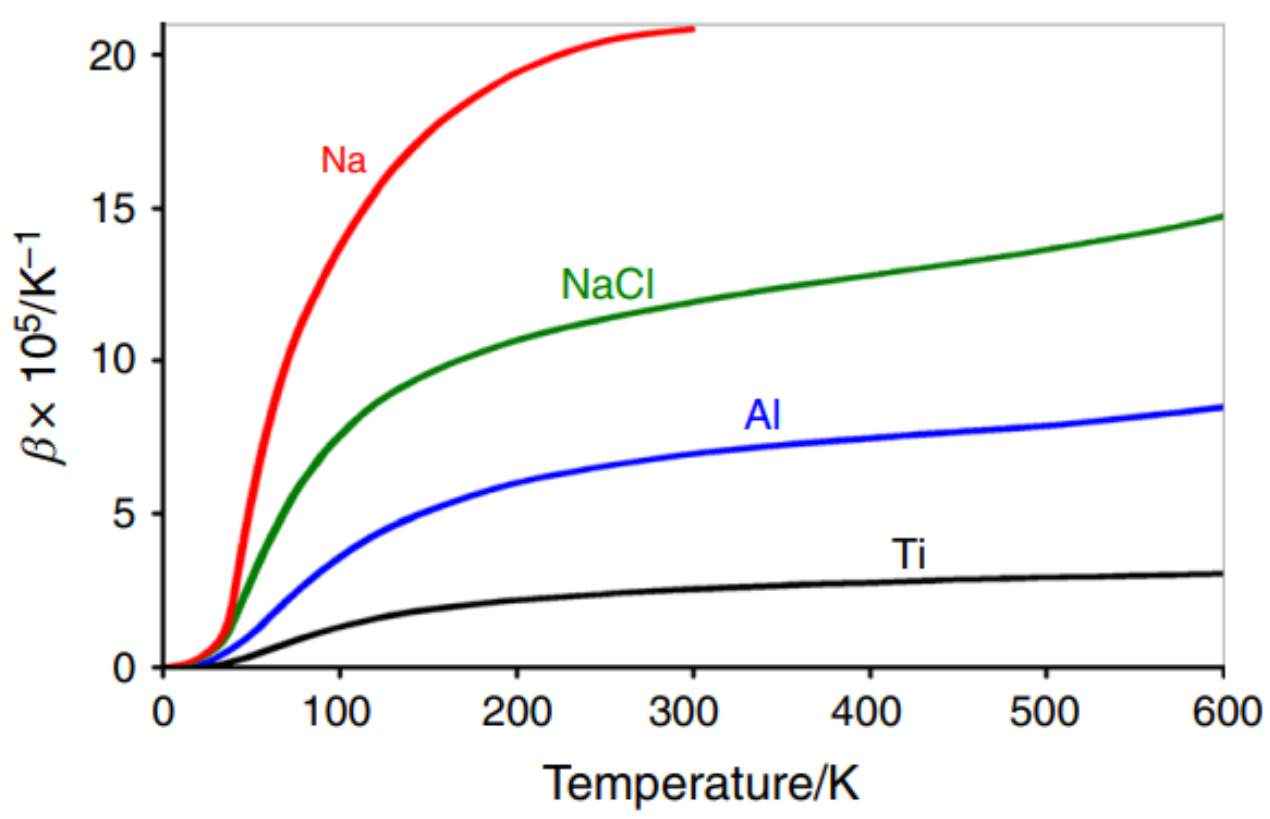

*4. Figure: Volumetric thermal expansion for several materials between 0 K and 600 K (Drebushchak, 2020)*

Figure 4. was created by Drebushchak based on values published in his work *Thermodynamics of Crystals*. Duane C. Wallace's book *Thermodynamics of Crystals* provides accurate descriptions of the thermal expansion coefficients of several materials at temperatures between 0 K and 600 K.

| ***T [K]*** | 25 | 50 | 75 | 100 | 125 | 150 | 175 | 200 | 300 |
|---|---|---|---|---|---|---|---|---|---|
| β ***[$10^{-5}K^{-1}$]*** | 0.40 | 2.93 | 5.70 | 7.58 | 8.78 | 8.60 | 10.20 | 10.68 | 11.95 |

*1. Table: Temperature-dependent values of* β *for NaCl (Wallace, 1972)*

The values listed in Table 1 show how the volumetric thermal expansion coefficient (β) of NaCl changes with temperature and serve as the basis for estimating the expected volume change under Martian surface conditions.

In our case, salt crystals that are relevant to the Martian environment should be evaluated. Although NaCl itself is unlikely to form stable liquid phases under present-day Martian humidity conditions, it provides a useful reference material because its temperature-dependent thermal expansion behaviour is comparatively well characterized in the literature. As an example, we calculated the volume change of NaCl due to temperature changes on Mars. We took the minimum temperature to be 150 K and the maximum temperature to be 300 K. Since the value of β for NaCl cannot be considered constant in the examined range, we used numerical integration on the points of the β(T) function and calculated the approximate relative volume change using the trapezoidal method, which gave us the following value:

$$\Delta V/V_0 = 0.0164$$

where ΔV is the difference between the final and the initial volumes, and $V_0$ is the initial volume. This dimensionless ratio corresponds to a relative volume change of 1.64%. I.e., the examined NaCl changes by 1.64% in this case between 150 and 300 K, i.e., between -123 and +27 $^{o}$C.

Accordingly, the temperature change occurring during an average Martian day results in an approximately 1.6% volumetric contraction of a NaCl crystal near the surface from afternoon (maximum temperature) to dawn (minimum temperature), followed by nearly the same amount of volumetric expansion from dawn to early afternoon.

Similar calculations can be performed for other salts relevant to Mars. The thermal expansion characteristics of gypsum (calcium sulfate dihydrate, $CaSO_4 \cdot 2H_2O$) were determined from unit cell volume data. According to the measurements, the volume of gypsum was recorded over a temperature range of 4.2 to 320 K, as shown in Figure 5. Figure 5 illustrates how the unit cell volume of gypsum changes with temperature, demonstrating the gradual thermal expansion of the mineral over the investigated range. In this case, the calculation was also performed for the 150–300 K temperature range. The unit cell volume of gypsum is 490.36 Å$^3$ at 150 K, while heating to 300 K increases it to 494.60 Å$^3$ (Schofield et al., 1996). Based on these values, the relative volume change between 150 and 300 K is 0.865%. Although gypsum is not among the most effective deliquescent salts under present-day Martian conditions, it provides a useful hydrated mineral reference for comparing the magnitude of thermally induced volume changes among different Mars-relevant evaporitic materials.

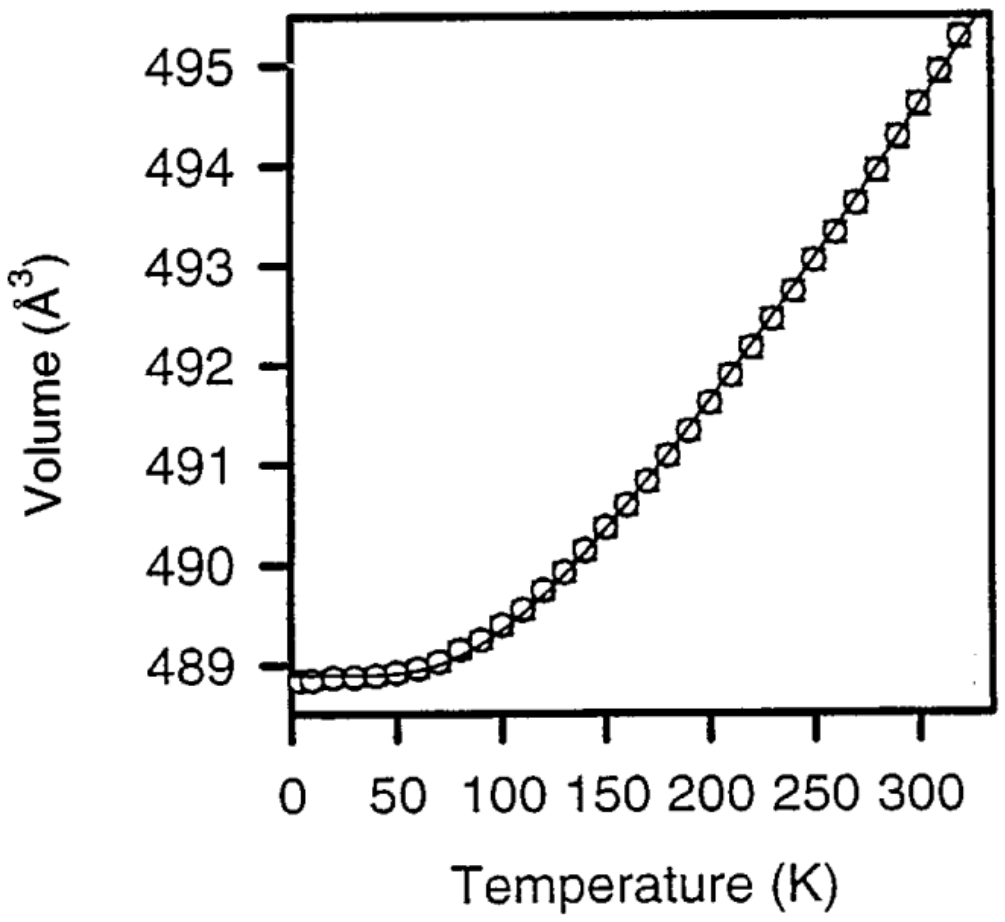

*5. Figure: Change in volume of calcium sulfate dihydrate as a function of temperature (Schofield et al., 1996)*

The α- and β-$MgSO_4$ are two different crystal structure modifications of anhydrous magnesium sulphate. Magnesium sulphates are considered here primarily as Mars-relevant evaporitic minerals and comparative mechanical analogues for evaluating thermally induced volume changes in different crystalline salt phases. They are chemically identical, but their structure, stability, and thermal expansion differ measurably. The temperature dependence of their volume is illustrated in Figure 6, which compares the thermal expansion behaviour of the two polymorphs. Their thermal expansion was calculated in a similar way to NaCl, and the volumetric thermal expansion data are as follows (Fortes et al., 2007):

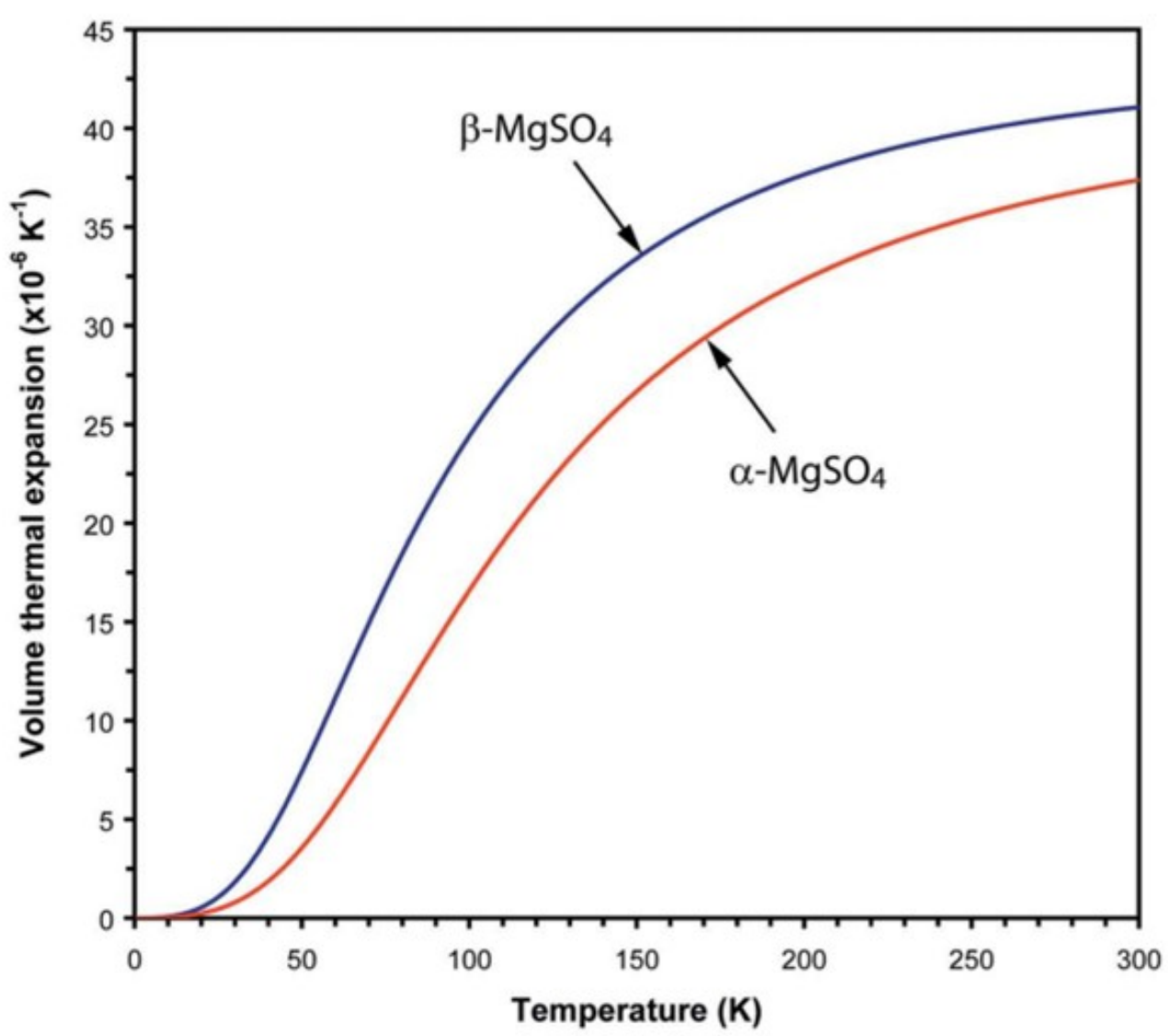

6. Figure: *Change in volume of α- and β-$MgSO_4$ as a function of temperature (Fortes et al., 2007)*

| ***T [K]*** | 150 | 200 | 250 | 300 |
|---|---|---|---|---|
| α-$MgSO_4$ : **β *[$10^{-6}K^{-1}$]*** | 26.65 | 32.47 | 35.54 | 37.39 |
| β-$MgSO_4$ : **β *[$10^{-6}K^{-1}$]*** | 33.49 | 37.78 | 39.89 | 41.04 |

*2. Table: Temperature-dependent values of* β *for* α- *and* β-*$MgSO_4$*

The values presented in Table 2 summarize the temperature-dependent volumetric thermal expansion coefficients used in the calculations and provide the basis for estimating the expected volume changes of the two $MgSO_4$ polymorphs under Martian temperature conditions. Using these data, we obtained a relative volume change of 0.50% for α-*$MgSO_4$* and 0.57% for β-*$MgSO_4$*.

Similar calculations can also be performed for liquid brines relevant to Mars. However, thermal expansion of liquid $CaCl_2$ solutions is physically different from the thermal expansion of solid salt crystals. In liquid brines, expansion may influence internal pore pressure or fluid redistribution, whereas in solid crystals thermal expansion may influence fracture aperture evolution. For a 40.9% calcium chloride solution, the coefficient of volumetric thermal expansion is 0.00046 1/K (The Engineering ToolBox, 2009), however unfortunately few Mars relevant salts have been measured in this aspect. Using the formula above, the relative volume change in this case is 6.9% for the same temperature range. If we have a 5.8% solution, this value decreases to 3.75% (The Engineering ToolBox, 2009). There are few reliable sources in the literature for accurate thermal expansion coefficients, and data for different types of salts are also incomplete.

These results demonstrate that thermally driven volume changes in both solid Mars-relevant salts and liquid brines may be non-negligible under Martian temperature conditions, including NaCl (1.64%), $CaCl_2$ solutions (3.75–6.9%), gypsum (0.87%), and $MgSO_4$ polymorphs (~0.5–0.6%). However, the mechanical consequences of thermal expansion likely differ between solid crystals, liquid solutions, and hydration-related phase or volume changes. Direct thermal expansion data for hydrated calcium perchlorates under Mars-relevant conditions remain limited; furthermore, calcium perchlorate hydrates possess substantially different crystal structures and hydration states compared to salts such as cubic NaCl or sulfate minerals considered here. Therefore, the presented calculations should be regarded as preliminary first-order estimates rather than quantitative predictions for $Ca(ClO_4)_2$ itself. Microbial colonization in porous environments commonly occurs in micrometer-scale pores and fissures, with frequently colonized pore diameters reported in the ~2.5–9 µm range (Jin and Sengupta, 2024). Therefore, thermally induced volume changes may become biologically relevant if they are capable of modifying crack apertures on similar micrometer scales.

## 3.1 Steps of the modelled process

The possible model for a habitable micro-environment on Mars is presented below supported by the previously listed aspects. Diurnal and longer-term temperature fluctuations under martian conditions may generate strains that produce or widen cracks, establishing contact between the interior of the crystal and the external environment, thus an internal cavity can come into contact with the atmosphere above the crystal. The following daily cycle might emerge and support the habitability of the crystal interior: as a result of nighttime cooling, atmospheric gases with increasing relative humidity may promote the condensation of atmospheric water vapor onto the surface of the hygroscopic crystal.

The measured DRH values depend not only on temperature but also to a large extent on the state of hydration and the stability of the crystal phase. Mars-relevant hygroscopic salts exhibit strongly composition- and temperature-dependent deliquescence relative humidity (DRH) values. Such variability is illustrated by DRH values reported for NaCl (DRH = 75% at 243 K; Gough et al., 2014), $NaClO_4 \cdot H_2O$ (DRH = 51% at 273 K and 64% at 228 K; Gough et al., 2011), $Mg(ClO_4)_2 \cdot 6H_2O$ (DRH = 42% at 273 K; Gough et al., 2011), $MgCl_2 \cdot 6H_2O$ (DRH = 33.7% at 273 K; Gough et al., 2014), and $CaCl_2$ hydrates, including $CaCl_2 \cdot 2H_2O$ (DRH = 12.9–19.5% between 223 and 273 K; Gough et al., 2016) and $CaCl_2 \cdot 6H_2O$ (DRH = 51.7–80.2% between 223 and 253 K; Gough et al., 2016). Calcium perchlorate ($Ca(ClO_4)_2$) may exhibit particularly low DRH values under certain hydration states and experimental conditions, with reported deliquescence thresholds ranging from ~5% up to ~55% depending on hydration state and temperature (Nuding et al., 2014). Calcium perchlorate ($Ca(ClO_4)_2$) exhibits particularly low DRH values, with reported deliquescence as low as ~5% at 273 K, although values may extend up to ~55% depending on hydration state (Nuding et al., 2014). Higher DRH values are observed for K-bearing salts, such as $KClO_4$ (DRH = 92% at 253 K; Gough et al., 2014) and KCl (DRH = 84% at 293 K; Gough et al., 2014).

The emerging microscopic liquid salty solution, as a layer, is expected to follow the dry surface and engulf the crystal together with its fracture surfaces, and enter deeper cavities along the opening crack, primarily as a microscopic film. Thanks to this process, the interior of a salt crystal can become wet at night through cracks that open up as a result of shrinkage. This process brings liquid into the crystal, however because of the low nighttime temperature possibility for metabolism might not emerge.

In the morning, when the temperature rises, the cracks close as the salt crystal begins to expand. Previous numerical studies suggest that thermally induced stresses under Martian conditions may be sufficient to support crack formation and subcritical crack growth (Eppes et al., 2015). However, crack closure does not necessarily imply complete sealing, and connected micro-scale pathways may still permit vapor escape under Martian daytime conditions, which substantially reduces the escape rate because of the highly limited size of microscopic voids still left there during daytime. Previous studies also suggest that even thin soil cover can substantially inhibit sublimation and vapor diffusion on Mars, while brines may further reduce evaporation due to their lower saturation vapor pressure (Martínez and Renno, 2013). Therefore, the proposed microenvironment should presently be regarded as a preliminary conceptual hypothesis requiring dedicated vapor-transport modeling. The internal wet surface that formed in the evening can keep the wetness inside, while the surface in contact with the atmosphere dries out due to the increasing temperature and related dry atmosphere. At this point, a hypothetical Martian organism can access water inside the salt crystal interior even in the warmer daytime conditions, which would not be possible on the open surface due to the generally severe dryness at that time. Based on these parameters, salt crystals may provide potential habitats for microorganisms considering the daily temperature and relative humidity changes. Figure 7 shows the changes in a daily cycle under favourable environmental conditions.

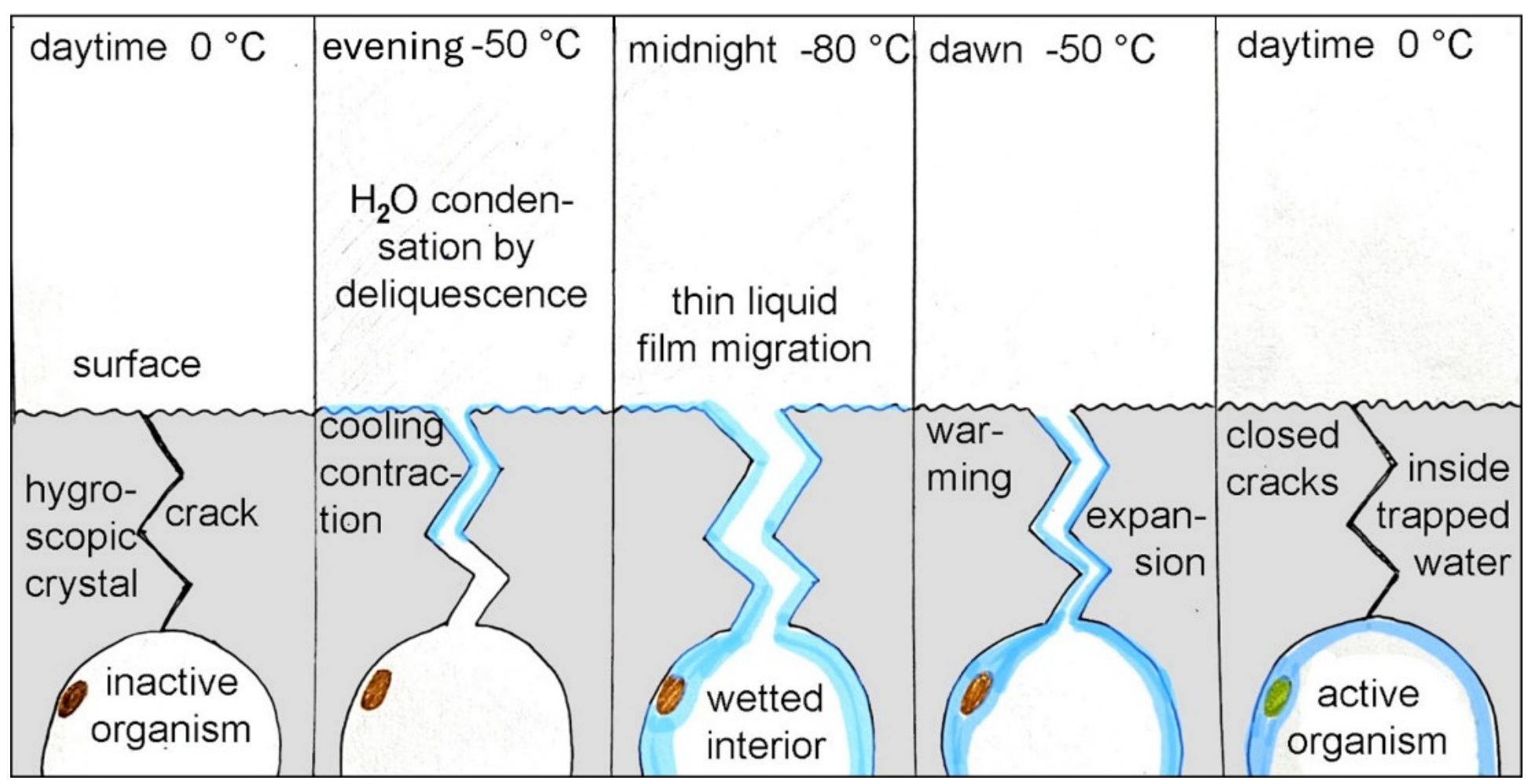

7. *Figure*: *Flowchart of a daily cycle.*

## 3.2. The possibility of deliquescence

In this section the areal distribution of model calculation based deliquescence (e.g. microscopic liquid water emergence) was evaluated, including the spatial correlation between the areal distribution of observed hygroscopic salts and locations favourable for the liquid water emergence. Because temperature-dependent thermal expansion data are comparatively well characterized for NaCl, it was used here primarily as an illustrative mechanical reference material, while hygroscopic salts such as $Ca(ClO_4)_2$ were considered in relation to present-day Martian deliquescence potential.

Although relative humidity near the surface of Mars can reach saturation levels at night and in the early morning, as evidenced by observations made by previous landing units, the combination of high relative humidity and sufficiently warm temperatures (Brass, 1980) required for NaCl deliquescence to occur is rarely met under current surface conditions on Mars. Therefore, it is unlikely that NaCl can exist in a liquid phase on the surface of Mars (Nguyen et al., 2024). NaCl is not expected to form a liquid phase on the surface of Mars, as experiments show that it only deliquesces at higher relative humidity levels, which are almost never reached on Mars due to its low humidity (Nguyen et al., 2024). In contrast, salts containing $Ca^{2+}$, such as $Ca(ClO_4)_2$, $CaCl_2$, and $MgCl_2$, are capable of binding water and forming brine even at lower relative humidity, thus potentially providing a liquid phase on the surface of Mars or in the subsurface environment (Nguyen et al., 2024). Experimental studies further suggest that $CaCl_2$- and $MgCl_2$-rich systems can enhance water retention and stabilize thin briny films under Mars-like conditions through ion-specific hydration effects and interactions with mineral substrates (Nguyen et al., 2024). Unfortunately, in the case of $Ca(ClO_4)_2$, which has been studied previously, no thermal expansion coefficient can be found that could be used to calculate the relative volume change.

Although solutions rich in sodium chloride do not pose a fundamental limitation for many microorganisms, salts containing calcium and magnesium, especially when combined with perchlorates, result in significantly lower water activity and thus provide a much narrower range of biological habitability (Stevens and Cockell, 2023).

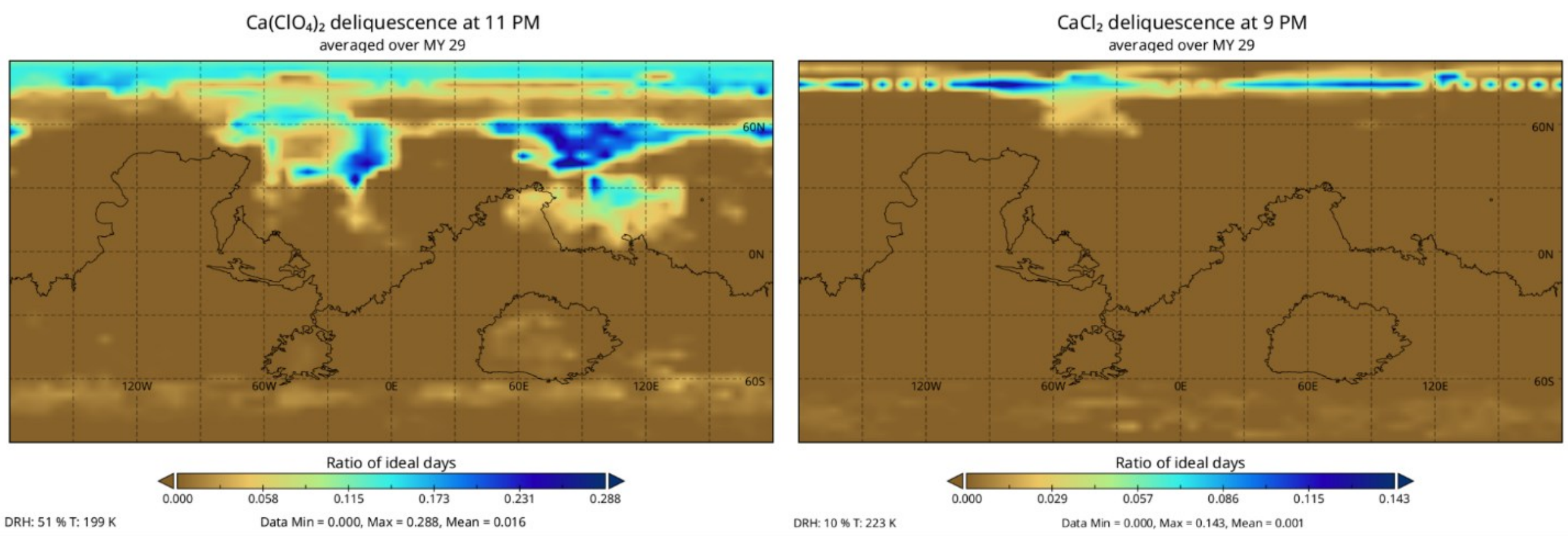

*8. Figure: Deliquescence possibility of $Ca(ClO_4)_2$ (left) and $CaCl_2$ (right) averaged over the 29$^{th}$ Martian year.*

In Figure 8. the possibility of deliquescence is shown for $Ca(ClO_4)_2$ (left) and $CaCl_2$ (right) over a full Martian year. The ratio of ideal days reflects the fraction of sols that exhibit favourable conditions at the specified local time; for example, a value of 0.1 at 9 PM indicates that on 10 % of the 669 sols, the value at 9 PM satisfied the deliquescence criteria. This would mean that on approximately 67 sols at 9 PM the T and RH conditions for deliquescence were met. Note, that this ratio does not include information of the seasonal distribution or whether the sols were consecutive. We refer readers interested in daily and seasonal trends to B. Pál et al., 2019 and Pál and Kereszturi, 2020 . Both salts exhibit a higher chance of deliquescence in the northern hemisphere. For $Ca(ClO_4)_2$ the regions around Acidalia Planitia, Utopia Planitia, Isidis Basin and some parts of the Hellas Basin show a higher possibility, while for $CaCl_2$ the possibly ideal region is concentrated around Acidalia Planitia. The regions showing higher possibility of deliquescence exhibit qualitatively similar spatial trends to the results presented by Rivera-Valentín et al. (2020), particularly near the edges of the polar ice caps.

However, it should be noted that our values represent the fraction of sols with favourable conditions at a given local time, and are therefore not directly comparable to studies reporting the fraction of favourable hours over a full Martian year. In both cases there is a higher chance at the edges of the ice caps that might be caused by the seasonal sublimation and subsequent recession of the polar ice caps. As $CO_2$ ice sublimates during local spring, the surface albedo decreases and temperatures rise, exposing the underlying $H_2O$ ice, which may locally increase atmospheric water vapor and create briefly favourable conditions for deliquescence (Pál & Kereszturi 2022). Gough et al. (2016) found that $CaCl_2$ has a remarkable ability to persist as a metastable, supersaturated brine, and once liquified it did not recrystallize until single-digit RH values were reached. This means that in the Martian subsurface $CaCl_2$ could exist in metastable form for over half a sol. Therefore, metastability alone may already prolong the persistence of liquid brines under Martian conditions, while partial crack closure could potentially further reduce vapor loss by limiting exchange with the surrounding atmosphere.

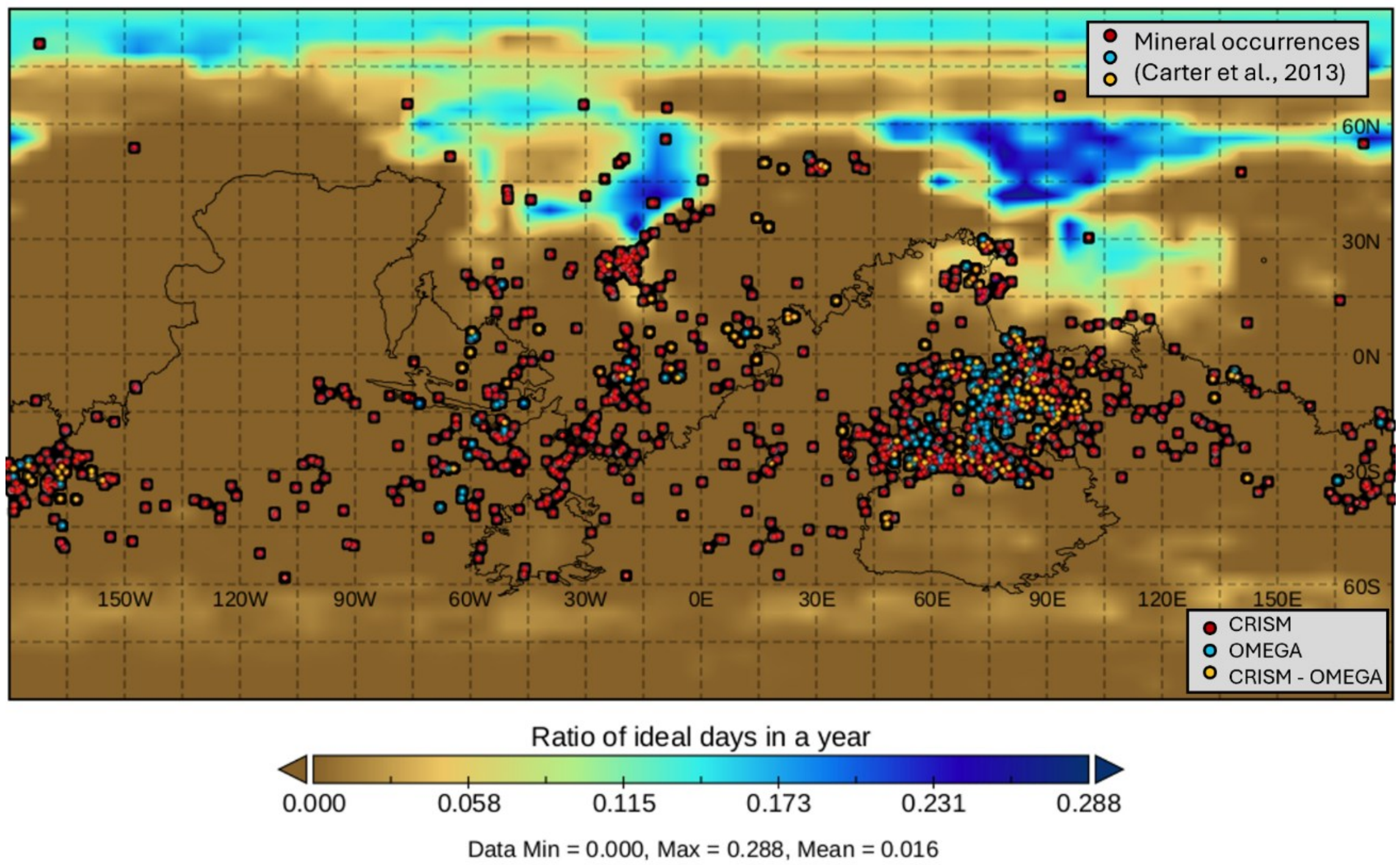

*9. Figure: Deliquescence potential of $Ca(ClO_4)_2$ with shades from brown toward blue, and locations of salt occurrences with red colour based Carter et al., 2013.*

Based on Figure 9., it can be concluded that salt crystals were detected at some locations (Carter et al., 2013) where the probability of deliquescence is higher (see the red dots in the bluish areas, especially at the top middle part of the map). When examining $Ca(ClO_4)_2$, the favourable area is located in the southern part of Acidalia Planitia. At approximately six sites in this region, the probability of deliquescence during a Martian year is about 17–20%, which corresponds to a cumulative of roughly 100–130 sols. This indicates that the process occurs often enough to be considered a reliable model approximation.

# 4. Discussion

The proposed mechanism should presently be regarded as a conceptual model requiring future quantitative retention modeling and experimental validation under Mars-analog conditions. Based on the results listed above, salt crystals can provide potential habitats for microorganisms. Due to the cold evening temperatures, cracks in a crystal may open up as a result of shrinkage, allowing any interior void to get connected with the ambient air. Thanks to the increase in relative humidity at night, the hygroscopic salt can liquefy and enter the deeper cavity. The process is supported by deliquescence, and the resulting microscopic liquid layer could easily migrate along the surfaces toward the interior of the crystal. Subsequently, when the temperature begins to rise in the morning, the volume of the crystal will increase, closing the crack that formed during the night, potentially before the inner cavity dries out completely, thus allowing a hypothetical organism to access liquid water at all times of the day inside the salt crystal. This process can thus provide protection against both daytime drying and UV radiation.

However, to go beyond the possibility outlined by the model, several unanswered questions exist in this topic, which can be answered with further research. These possibilities are discussed below, partly in order to orient related next activities, and how the model could be verified.

## 4.1. Evaluation of further conditions and possibilities

Further theoretical argumentation is discussed below to better estimate the possibility of how the above presented model might work on Mars.

The thermal expansion calculated from the values shown in the Results section is significant in both cases, ranging between 150 K and 300 K. Thus this value is not a negligible factor in our case. Unfortunately, the literature is lacking in thermal expansion coefficients and their temperature variations, but the measured NaCl based estimated values may be sufficient as a starting point. It would be important to determine further values in order to establish whether this conclusion can be drawn for all Mars-relevant salts, namely that they have significant thermal expansion, since thermoelastic stress generation and crack evolution may additionally depend on mineralogical and microstructural properties (Eppes et al., 2015). However, crack closure alone does not necessarily imply complete sealing against vapor loss, and dedicated vapor-transport modeling would be required to evaluate the long-term stability of such hypothetical microenvironments.

In addition to simple thermal expansion, several other processes may influence crack evolution in hygroscopic salts under Martian conditions, including hydration/dehydration-driven volume changes (preferable minerals are low water uptake once which shrink during cooling), deliquescence (preferable such minerals which trap and transport the water molecules on their surface), efflorescence, dissolution, recrystallization, and crystallization pressure (however the last three effects are expected damped during the cold nights). These processes may either enhance or counteract thermally induced crack opening and closure and were not explicitly modelled in the present study. These processes should be evaluated in future laboratory experiments under Mars-relevant environmental conditions.

In the case of individual salts, the deliquescence process is more likely to occur in the northern parts, as shown in Figure 8. below. Comparing where hygroscopic salts have been detected, it can be observed that the southern part of Acidalia Planitia and the Isidis Basin area have the most favourable values. Although smaller, the Hellas Basin area has a non-negligible probability, with its northern edge having the densest salt occurrence (Carter et al., 2013). Based on the results, it can be concluded that salt crystals occur in several places in areas where the probability of deliquescence is highest. In the case of $Ca(ClO_4)_2$, the area of interest lies in the southern sector of Acidalia Planitia in the high northern latitude regions.

In addition to the parameters discussed so far, the harmful effects of UV radiation must also be taken into account. The atmosphere on Mars does not provide as much protection against UV radiation as the Earth’s atmosphere, as the ozone layer there is much thinner. Therefore, the Sun's UV radiation can have a devastating effect on any microorganisms that may be found on the surface (Godin et al., 2019), however, the UV-shielding and light-transmission properties of salt-rich microenvironments likely depend strongly on mineral composition, transparency, dust content, and substrate structure. Transparent halite crystals may permit partial light penetration, whereas perchlorate-, sulfate-, or dust-rich substrates may provide stronger UV shielding but lower visible-light transmission.

In 2015, Schuerger studied the survival of two bacterial species in a Mars Simulation Chamber (MSC). The measurements took into account low pressure, UV radiation (UVA, UVB, UVC), the composition of the Martian atmosphere, as well as temperature and dust concentration at different optical depths. Schuerger's measurements show that if we want a habitable environment, we need an extra protective layer against UV radiation. Most of the bacteria he studied became unviable within a few seconds, and within 60 min, the living cells had completely disappeared. For UV shielding salty solutions, ice layers, dust, or even thin rocks can be used as radiation shielding while still allowing some light through for photosynthesis, but significantly reducing UV radiation.

In addition to their shielding role, hygroscopic salts may also provide transient water availability through deliquescence–efflorescence cycles. As pointed out by Schulze-Makuch (2024), microorganisms adapted to hyperarid environments could survive by utilizing very small amounts of water present only for short periods within salt matrices. This interpretation is consistent with our proposed mechanism and may also help to explain the ambiguous results of the Viking life-detection experiments, where signs of activity appeared after wetting, but no clear organic compounds were detected (Schulze-Makuch, 2024).

Accordingly, an optimal zone may occur at shallow depths, especially in transparent materials (e.g., gypsum, rock salt, ice). Marschall et al. (2012) examined the UV filtering capacity of the surface. They found that a 1-2 mm thick layer of Mars-simulating material (JSC-1 and basalt sand) provided sufficient protection against harmful radiation. The depth to which visible light may still penetrate sufficiently for potential photosynthetic activity is approximately 4-6 mm in the Mars simulant granular materials JSC-1 (<1 mm particle size) and basaltic sand (63 µm – 1.5 mm particle size). Combining these two parameters, there is a zone close to the surface where hypothetical microorganisms may be able to photosynthesize if other conditions are favourable (Marschall et al., 2012). Laboratory tests also found that certain cyanobacterial pigments can still be detected after long exposure to Mars-like UV radiation (Baqué et al., 2015). However, the actual photon flux reaching partially closed cracks or internal cavities would likely depend strongly on crack geometry, mineral transparency, dust coverage, and optical scattering effects, and therefore would require dedicated optical modeling.

Mickol et al. (2017) placed bacteria in different solutions (sterile distilled water, 5% and 10% magnesium sulphate ($MgSO_4$) solutions) and examined separately how they reacted to the solution being in a liquid or frozen state (-25°C). They observed that after 10 s of UV exposure, the number of surviving cells decreased by 3-4 orders of magnitude, while after 60 min of exposure, all living cells disappeared. Similar results were observed in the case of frozen salt solutions. Although these solutions reduced UV intensity by 12%, they did not provide significant protection for the survival of life (Mickol et al., 2017). Such liquid solutions might emerge as a microscopic thin surface layer on hygroscopic salts. Subsequently, they examined how thin layers of dust and rock affect the survival of life despite strong radiation. The bacteria were first placed in a 5% $MgSO_4$ solution and then frozen. The resulting ice layer was covered with two different types of Martian analogue soil, such as Hawaiian palagonite, which is an iron-rich volcanic rock, and crushed basalt. After 60 min of UV radiation, it was observed that the two protective layers provided similar protection. The tests showed that a dust layer approximately 3-4 mm thick completely blocks harmful radiation. Based on these measurements, a shallow layer beneath the Martian surface may be the best place for potential life to survive (Mickol et al., 2017).

A greater problem is the ionizing particle radiation from the solar wind and galactic cosmic rays. Against these agents on the scale of a meter thick regolith (or other solid material) layer gives enough shielding, and the millimetre thin mineral cover is not enough here. However, in theory, highly adapted organisms (Olsson-Francis and Cockell, 2009) could repair their internal radiation-driven damages in the active phase, the question is: could an organism provide enough regular repair using the short periods useful for metabolism and related living activities under the harsh conditions. Based on our current knowledge this cannot be properly answered (Michalski et al., 2018), although it cannot be excluded; however, it remains a highly theoretical question.

Even if the given micro-environment might be habitable, it does not necessarily mean it is occupied by organisms (Cockell et al., 2012). However, if life once emerged on early Mars, prolonged exposure to harsh environmental conditions may have allowed the evolution of adaptation strategies supporting survival in localized microenvironments.

In addition to the environmental challenges mentioned above, it is important to consider several other factors. Laboratory experiments have shown that bacteria may be capable of growth even under extremely low atmospheric pressure (Schuerger and Nicholson, 2016). Their results indicate that multiple bacterial species can grow at pressures typical of Mars, only 0.7 kPa, in a $CO_2$-rich, oxygen-free atmosphere and at low temperatures. Although growth was slow and required a stable supply of water and nutrients, pressure alone does not appear to be the primary obstacle to life (Schuerger and Nicholson, 2016). These experimental conditions differ from the transient and periodically wetted microenvironments proposed here for present-day Mars, where liquid water availability would likely be spatially and temporally limited.
The transient presence of liquid water alone is not sufficient to indicate biological habitability on present-day Mars. Habitability is also limited by the physical and chemical properties of the brines. An important parameter is water activity ($a_w$), which describes the amount of biologically available water. Rivera-Valentín et al. (2020) showed that metastable Martian brines generally do not simultaneously reach the temperature and water activity conditions required for known terrestrial life. Potentially habitable conditions would require at least $a_w > 0.5–0.6$ and $T > 250–255$ K, while Martian brines commonly form below 225 K. Our model supports the existence of water at elevated temperature e.g. under conditions of elevated water activity. .
In addition, water activity alone is not sufficient to determine habitability. Heinz et al. (2021) showed that high ionic strength, chaotropicity, and destabilization of proteins and membranes may further limit microbial growth. The halotolerant yeast Debaryomyces hansenii showed growth only down to about $a_w \approx 0.83$. Perchlorates may be especially restrictive because they both lower water activity and chemically stress biomolecules. However, some experiments suggest that small amounts of water produced by deliquescence may temporarily reactivate microbial metabolism. Maus et al. (2020) observed methane production by some methanogenic archaea under Mars-analog conditions after deliquescence-driven wetting. These results suggest that present-day Martian brines may allow temporary survival or limited metabolic activity in some cases, but long-term habitability remains highly uncertain. Our model supports the extended presence of such brines. In contrast, field studies in the Atacama Desert have demonstrated that biomolecules such as ATP and cellular pigments degrade extremely rapidly, even when partially shielded by salt crystals or mineral grains (Arens et al., 2025). This suggests that even if transient salt-crystal microhabitats could temporarily stabilize liquid water on Mars, long-term survival and biosignature preservation may remain strongly limited by radiation-driven degradation processes.

Despite this, Martian rock samples regularly contain detectable organic matter. Investigations by the Perseverance rover identified organic-carbon–bearing mudstones within the Bright Angel formation in Jezero Crater. These rocks contain microscopic mineral nodules composed of iron phosphate and iron sulphide. Measurements suggest that the organic material may have become bound to these minerals through post-depositional redox processes, enabling its long-term preservation. Such mineral traps can effectively stabilize organic molecules, even if they would rapidly degrade on the surface (Hurowitz et al., 2025).

Overall, this creates a paradox: while laboratory and field studies show that the present-day Martian surface rapidly destroys organic matter and is highly unfavourable for life, traces of organics are nevertheless preserved in Martian rocks. The presented model concerns the current possibility of a unique environment type and the inferred processes there, which supports a focused analysis of a realistic candidate habitable environment in the future (see below).

## 4.2. Potential terrestrial analogies

Studying similar astrobiologically relevant water trapping within hygroscopic salts under terrestrial conditions can help us understand the behaviour of salts on the surface of Mars. Although it is very difficult to accurately model the extreme Martian environment in a laboratory, such as low air pressure, significant daily temperature fluctuations, and especially humidity control in a low-density small volume gas medium, there are places on Earth that can serve as potential and partial analogies providing an opportunity to gain more details of the behaviour of Martian salt crystals with embedded organisms. One such environment is at the Atacama Desert in Chile, which is known as one of the driest places on Earth, stretching approximately 1,000 km along the Pacific coast of South America. Exceptionally dry conditions can be found in the southern part of the desert, where the extreme dry conditions have remained virtually unchanged for 10-15 million years, making it one of the oldest deserts on Earth. The annual rainfall is typically less than 1 mm (Mckay et al., 2003). In the Yungay region of the desert, the barren landscape is dotted with small stones, some of which can be seen on the ground and others embedded in the soil. These stones provide a habitat for hypolithic cyanobacteria and algae (Casero et al., 2021). Their survival strategies can help us understand how to survive in similar Martian conditions. In such a barren, extreme climate desert, these microorganisms photosynthesize, and bacteria hiding in the vicinity of the stones can obtain moisture with the help of the microenvironment (McKay et al., 2003).

In the driest areas of the Atacama Desert, living organisms are found primarily within salt-rich substrates that are capable of absorbing water from atmospheric humidity (Davila and Schulze-Makuch, 2016). Davila and Schulze-Makuch (2016) and Schulze-Makuch (2016) show that, due to increasing aridity, habitats change in a characteristic sequence: soil-bound microbial communities are first replaced by rock-dwelling life forms, and then, under the most extreme conditions, life is restricted to hygroscopic, salt-rich environments. In this case, these salt-based microhabitats represent the last surface-level refuges for life in extremely arid environments. The microhabitats we propose fit into this final category, as they are also based on hygroscopic materials and periodically accessible moisture from the air.

Studies have reported halite (NaCl) minerals in the desert. Laboratory experiments have confirmed that halite has a strong moisture-binding capacity and aids the metabolism of associated cyanobacteria (Davila et al., 2013). Gómez-Silva et al. (2019) also investigated the possibility of life in the Atacama Desert, despite the incredible dryness. Following the experiences described by McKay et al. in 2003, the 2019 study summarized the results achieved over the past two decades, according to which the soils, sediments, and rocks found in this location contain a rich microbiota, such as archaea, bacteria, fungi, protozoa, algae, and viruses.

The rocks found in the desert may play a key role in sustaining life, as they serve multiple functions. They can act as water collectors and reservoirs, providing protection against significant temperature fluctuations, and can also reduce the impact of high ultraviolet radiation. Halite crystals, also known as rock salt, are commonly found in the Atacama Desert, and many rocks consist of more than 95% NaCl. These are evaporitic sedimentary rocks formed and left behind by evaporation. Halites serve as a home for microorganisms that are most adapted to high salt concentrations. Halite crystals, which are typically white or colourless, are partially transparent to sunlight. This may permit limited light penetration and potentially support photosynthetic microbial activity inside them (Gómez-Silva et al., 2019).

Analyses have shown that the microbial communities found in halite crystals also play a role in the carbon, nitrogen, and sulfur cycles, which are important in exploring the possibility of potential Martian life, as similar processes may play a role in the survival of a Martian microbial ecosystem (Gómez-Silva et al., 2019). This endolithic communities inside halite crusts with cyanobacteria within pore spaces (Wierzchos et al., 2006 and de los Ríos et al., 2010) uses deliquescence for water access enough for photosynthetic activity (Osano et al., 2014, Davila et al., 2010). Among model organisms and survival strategies Halobacteria (DasSarma, 2006) were suggested with elevated salt tolerance, as well as xerophiles (Grant, 2004) requiring small amounts of water. Studies on Atacama region extremophiles (Wierzchos et al., 2006 and de los Ríos et al., 2010) indicate they generally benefit from the hygroscopic minerals' water trapping abilities there (DiRuggiero et al., 2013) supporting colonizing bacterial communities inside various salty rocks (Davila et al., 2011).

Slank et al. (2024) also examined the Atacama Desert as a potential terrestrial analogue in their study. With this in mind, they conducted research in which they examined several Mars-relevant salts in the desert. The samples to be examined were placed in plexiglas boxes with small holes drilled in the lids to ensure interaction with atmospheric humidity. Several sensors were placed in each box to continuously measure temperature and relative humidity. The HOBO sensor indicated the presence of liquid water based on conductivity. The results showed that deliquescence can occur even in a dry, natural environment, but only in the case of salts with a low eutectic point, such as calcium perchlorate (Slank et al., 2024).

Sodium sulfate ($Na_2SO_4$, thenardite) is a desert salt mineral that undergoes volumetric changes due to daily temperature fluctuations, contributing to salt weathering processes (Cooke & Smalley, 1968). Since sulfate minerals are also present on Mars, $Na_2SO_4$ could potentially play a role in thermally induced rock weathering processes in Martian saline sedimentary environments.

# 5. Suggestions for future research

There are still questions and measurement gaps that hinder understanding and require further investigation in several areas. It would be worthwhile to investigate the phenomenon of liquefaction in the future for a variety of minerals and mixtures thereof, under conditions similar to those found on Mars. With this knowledge, we could learn about the behaviour of salt minerals that can provide suitable conditions for bacteria in the possible habitats described in this paper.

Testing the rates of crack closure during daytime warming and interior drying under Martian conditions will be necessary to evaluate the viability of the proposed microhabitat mechanism. In this context, it would be important to know the rate at which cracks that opened during the night-time cooling close during daytime warming and, in parallel, how quickly the interior of the mineral can dry out, since maintaining a potential habitat requires that it does not dry out during the day.

Another interesting parameter that requires further research is whether these hygroscopic salt minerals are capable of shrinking as a result of drying. More specifically, how effective is the change in crystal volume caused by shrinkage due to drying and expansion not only from temperature increase but also simultaneously from hydration changes, and what are the consequences for the cracks. Future studies should also distinguish between the effects of solid-state thermal expansion, liquid brine expansion, hydration/dehydration-driven volume changes, and crystallization/dissolution processes, since these mechanisms may influence crack evolution in fundamentally different ways. In addition, processes such as deliquescence, efflorescence, recrystallization, and crystallization pressure may either enhance or counteract thermally induced crack opening and closure under Martian conditions. A major limitation of the present study is the lack of reliable temperature-dependent thermal expansion data for hydrated calcium perchlorates and several other Mars-relevant hygroscopic salts under Mars-like conditions. Furthermore, very few thermal expansion coefficients have been accurately measured as a function of temperature in the literature. For a more detailed study, it would be important to have the thermal expansion coefficients of several Mars-relevant salts within the calculated temperature range.

Future studies should also distinguish between the effects of solid-state thermal expansion, liquid brine expansion, hydration/dehydration-driven volume changes, and crystallization/dissolution processes, since these mechanisms may influence crack evolution in fundamentally different ways.

It would also be worthwhile to further investigate whether bulk thermal expansion under Martian conditions can realistically result in crack opening and closing, since crack evolution likely depends on additional factors such as grain size, elastic properties, pre-existing microfractures and thermal gradients. Future studies should also examine whether crack closure under Martian conditions can sufficiently limit vapor transport to enable temporary liquid retention.

With further experiments, we can create a more accurate model of the processes on the surface of Mars and the potential biological possibilities associated with them. Currently, laboratory tests have confirmed the volume changes; however, a better understanding of hydration-water variations is still required. Water trapping by deliquescence is expected to operate under Martian conditions. Light penetration sufficient for photosynthesis is also expected and could be readily verified experimentally for specific minerals. Overall, the model is ready for future testing.

# 6. Conclusions

In this work, we present and evaluate a model that explores the formation of liquid water on Mars for hypothetical microorganisms during otherwise dry daytime conditions, facilitated by hygroscopic salts. Different Mars-relevant salts—including NaCl, gypsum, $MgSO_4$ polymorphs, and calcium salts such as $CaCl_2$ and $Ca(ClO_4)_2$—were considered from complementary perspectives to capture their diverse physicochemical behaviors. However, the lack of direct thermo-physical data for hydrated calcium perchlorates remains an important limitation, and therefore the proposed behaviour of $Ca(ClO_4)_2$ under Martian conditions should presently be regarded as preliminary. The dryness that characterizes the surface of Mars poses significant problems for the emergence of microbial life. A solution to this problem may be provided by hygroscopic salts, which have the ability to absorb atmospheric water vapor and to form a liquid solution (Gough et al., 2016, Martín-Torres et al., 2015, and Zorzano et al., 2009). However, for this to happen, it is important that the relative humidity exceeds the DRH value characteristic of the salt in question. Based on the measured values, such a salty solution can form for several hours in the morning and evening on the surface of Mars (Pál and Kereszturi, 2017). In the case of $CaCl_2$ and $Ca(ClO_4)_2$, which were examined separately, the process is more likely to occur in the northern hemisphere, and possibly at the edges of polar ice caps, which may be caused by the seasonal sublimation and subsequent retreat of the ice caps. However, it is important to note that biological activity depends not only on salt concentration but also on salt composition, as solutions rich in sodium chloride may be significantly less restrictive than environments dominated by calcium- or magnesium-containing salts. In particular, perchlorate-rich brines may exhibit high ionic strength and chaotropic effects that can destabilize biomolecules and strongly limit microbial growth under present-day Martian conditions (Heinz et al., 2021).

Various measurements have shown that salt crystals occur in many places on Mars (Carter et al., 2013 and Osterloo et al., 2008). These silicate, sulphate, and chloride deposits are found in more than 990 different locations on the surface, in craters, channels, or spots, especially in the equatorial region.
In the equatorial region, the daily temperature fluctuation of the salt crystals examined is significant, with a daytime maximum of 273 K and a nighttime minimum of 150 K, causing the material to undergo significant volume changes.
Based on the temperature-dependent volumetric thermal expansion coefficient β(T) reported by Drebushchak (2020) and Wallace (1972), the relative volume change of a NaCl crystal between 150 K and 300 K was calculated by numerical integration of β(T) over the temperature interval, yielding $\Delta V/V_0 = 0.0164$ (1.64%). Applying the same approach, or available literature values where β can be approximated as constant, we obtained relative volume changes of 3.75–6.9% for $CaCl_2$ solutions (depending on concentration), 0.87% for gypsum (Schofield et al., 1996), and approximately 0.50–0.57% for α- and β-$MgSO_4$ (Fortes et al., 2007) over the same temperature range. However, NaCl cannot form a liquid phase under Martian conditions due to low relative humidity (Nguyen et al., 2024). In contrast, hygroscopic salts such as $Ca(ClO_4)_2$ and $CaCl_2$ may remain more relevant under present-day Martian conditions. Nevertheless, direct thermal expansion data for hydrated calcium perchlorates remain very limited, and calcium perchlorate hydrates possess substantially different crystal structures and hydration states compared to cubic NaCl and sulfate minerals. Therefore, the presented calculations should be regarded only as preliminary first-order estimates rather than quantitative predictions for $Ca(ClO_4)_2$ itself.

However, the increase in volume during the daytime can close any cracks, which can then open up again in the cooler night due to shrinkage (Eppes et al., 2015).
The absolute amount of atmospheric water vapor reaches its maximum in the morning and afternoon hours, as this is when the Sun heats the surface, allowing moisture to escape in the form of water vapor. In the cooler evening hours, the regolith can absorb more water again (Titov, 2002). However, the relative humidity, according to measurements taken by the Mars Science Laboratory onboard the Curiosity rover, during daylight hours is very low, often approaching zero. At night, when temperatures reach the daily minimum, relative humidity rises and may approach saturation. However, high relative humidity is not accompanied by high absolute amount of atmospheric water vapor, as it fluctuated between only 30-75 ppm (Harri et al., 2014).

In addition to the parameters mentioned so far, the harmful effects of UV radiation must also be taken into account, as a significant proportion of the bacteria studied become unviable within a few seconds, and within 60 min, living cells disappear completely under Martian conditions (Mickol et al., 2017 and Schuerger, 2015). It is also probable that some hygroscopic minerals on Mars may provide partial UV shielding at depths of 2–3 mm while, depending on mineral composition and transparency, still transmitting limited visible light (Marschall et al., 2012). Therefore, although such shallow crystalline microenvironments may temporarily improve local water stability and UV protection, they would likely provide only limited shielding against ionizing particle radiation under present-day Martian conditions. Consequently, ionizing radiation remains a major unresolved limitation for the long-term habitability of such near-surface environments.

# 8. Acknowledgements

AB acknowledges the support on her BSc thesis for the Konkoly Thege Miklós Astronomical Institute (CSFK, HUN-REN) what served as background for this work as well to the internal consultant, Krisztina Éva Szentirmayné Gabányi, for supporting the creation of this paper.

## Data availability statement

Raw data were generated at the LMDZ Laboratory (Laboratoire de Météorologie Dynamique) at Sorbonne Université. Derived data supporting the findings of this study are available from the corresponding author, following permission from the LMDZ Laboratory, on request. The codes used in this study are currently available from B. D. Pál, upon reasonable request.

# 9. References

- Angell, P., Werner, H. and Christensen, P.R. *Diurnal and seasonal temperature variations along the Mars Curiosity rover traverse: A comparison of THEMIS remote sensing data with ground truth.* 51st Lunar and Planetary Science Conference, 2020.

- Arens, F.L., et al., 2025. Biomolecular degradation in a Mars-analog environment: rapid decay of ATP and pigments in the Atacama Desert. Sci. Rep. doi: 10.1038/s41598-025-16197-w.

- Baqué, M., Verseux, C., Böttger, U., Rabbow, E., de Vera, J.-P. and Billi, D. *Preservation of biomarkers from cyanobacteria mixed with Mars like regolith under simulated Martian atmosphere and UV flux.* Astrobiology, 15:667–682, 2015, doi: 10.1007/s11084-015-9467-9

- Becker, G.F. and Day, A.L. *The linear force of growing crystals.* Proceedings of the Washington Academy of Sciences, 7:283–288, 1905.

- Brass, G.W. *Stability of brines on Mars.* Icarus, 42:20–28, 1980, doi: 10.1016/0019-1035(80)90237-7.

- Carter, J., Poulet, F., Bibring, J.-P., Mangold, N. and Murchie, S. *Hydrous minerals on Mars as seen by the CRISM and OMEGA imaging spectrometers: updated global view .* Journal of Geophysical Research: Planets, 118, 831–858, 2013.

- Casero, M.C., Meslier, V., DiRuggiero, J., Quesada, A., Ascaso, C., Artieda, O., Kowaluk, T. and Wierzchos, J. *The composition of endolithic communities in gypcrete is determined by the specific microhabitat architecture.* Biogeosciences, 18(3):993–1007, 2021, doi: 10.5194/bg-18-993-2021.

- Cockell, C.S., Balme, M., Bridges, J.C., Davila, A. és Schwenzer, S.P. *Uninhabited habitats on Mars.* Icarus, 217:184–193, 2012, doi: 10.1016/j.icarus.2011.10.025.

- Cooke, R.U., & Smalley, I.J. *Salt weathering in deserts.* Nature, 220(5173), 1226–1227, 1968.
- DasSarma, S*. Extreme halophiles are models for astrobiology.* Microbe, 2006, 1(3), 120–126.
- Davila, A.F., Duport, L.G., Melchiorri, R., Jänchen, J., Valea, S., de los Ríos, A., Fairén, A.G., Möhlmann, D., McKay, C.P., Ascaso, C. and Wierzchos, J. *Hygroscopic salts and the potential for life on Mars.* Astrobiology, 10, 2010.
- Davila, A.F., Hawes, I., Ascaso, C. and Wierzchos, J. *Salt deliquescence drives photosynthesis in the hyperarid Atacama Desert.* Microbiology, 5:583–587, 2013, doi: 10.1111/1758-2229.12050.
- Davila, A.F. and McKay, C.P. *Salt flats in Terra Sirenum: a site to search for extant and extinct life on Mars.* Analogue Sites for Mars Missions, 6028, 2011.
- Davila, A.F. and Schulze-Makuch, D. *The last possible outposts for life on Mars.* Astrobiology, 16:159–168, 2016, doi: 10.1089/ast.2015.1380
- de los Ríos, A., Valea, S., Ascaso, C., Davila, A.F., Kastovsky, J., McKay, C.P., Gómez-Silva, B., Wierzchos, J. *Comparative analysis of the microbial communities inhabiting halite evaporites of the Atacama Desert.* International Microbiology, 13:79–89, 2010, doi: 10.2436/20.1501.01.113.
- DiRuggiero, J., Wierzchos, J., Robinson, C.K., Souterre, T., Ravel, J., Artieda, O., Souza-Egipsy, V., and Ascaso, C. *Microbial colonisation of chasmoendolithic habitats in the hyper-arid zone of the Atacama Desert.* Biogeosciences, 10:2439–2450, 2013, doi: 10.5194/bg-10-2439-2013.
- Drebushchak, V.A., 2020. *Thermal expansion of solids: review on theories.* J. Therm. Anal. Calorim. doi: 10.1007/s10973-020-09370-y.
- Eppes, M.-C., Willis, A., Molaro, J., Abernathy, S. and Zhou, B. *Cracks in Martian boulders exhibit preferred orientations that point to solar-induced thermal stress.* Nature Communications, 6:6712, 2015, doi: 10.1038/ncomms7712.
- Fauré, B., Beck, P., Schmitt, B., Pommerol, A., Thomas, N. and Poch, O. Water adsorption and deliquescence processes on Mars-relevant salts explored by molecular dynamics simulations. ACS Earth Space Chem. , 2023, doi: 10.1021/acsearthspacechem.3c00072.
- Fletcher, R.C. and Merino, E. *Mineral growth in rocks: kinetic-rheological models of replacement, vein formation, and syntectonic crystallization.* Geochim. Cosmochim. Acta 65, 3733–3748. doi:10.1016/S0016-7037(01)00726-8.
- Forget, F., Hourdin, F., Fournier, R., Hourdin, C., Talagrand, O., Collins, M., Lewis, S.R., Read, P.L. and Huot, J.-P. *Improved general circulation models of the Martian atmosphere* *from the surface to above 80 km.* Journal of Geophysical Research: Planets, 104(E10):24155–24176, 1999, doi: 10.1029/1999JE001025.

- Fortes, A.D., Wood, I.G., Vočadlo, L., Brand, H.E.A., Knight, K.S., 2007. Crystal structures and thermal expansion of α-MgSO4 and β-MgSO4 from 4.2 to 300 K by neutron powder diffraction. J. Appl. Crystallogr. 40 (4), 761–770. doi:10.1107/S0021889807029937.

- Godin, P.J., Stone, H., Bahrami, R., Schuerger, A.C. and Moores, J.E. *Habitability of bodies of water on ancient Mars: Attenuation of UV radiation from aqueous solutions of minerals found on Mars.* 50th Lunar and Planetary Science Conference (LPI Contrib. No. 2132), 2019.

- Goff, J.A., Gratch, S., 1946. *Low-pressure properties of water from-160 to 212F.* Trans. ASHVE 95-122.

- Gómez-Silva, B., Vilo-Muñoz, C., Galetovic, A., Dong, Q., Castelán-Sánchez, H.G., Pérez-Llano, Y., Sánchez-Carbente, M.R., Dávila-Ramos, S., Cortés-López, N.G., Martínez-Ávila, L., Dobson, A.D.W. and Batista-García, R.A. *Metagenomics of Atacama lithobiontic extremophile life unveils highlights on fungal communities, biogeochemical cycles and carbohydrate-active enzymes*. Article, 2019, doi: 10.3390/microorganisms7120606

- Gough, R.V., Chevrier, V.F., Baustian, K.J., Wise, M.E., Tolbert, M.A. *Laboratory studies of perchlorate phase transitions: support for metastable aqueous perchlorate solutions on Mars. Earth Planet. Sci. Lett.* 2011, doi: 10.1016/j.epsl.2011.10.026.

- Gough, R.V., Chevrier, V. and Tolbert, M.A. *Formation of aqueous solutions on Mars via deliquescence of chloride-perchlorate binary mixtures.* Planet. Space Sci., 393:73-82, 2014, doi: 10.1016/j.epsl.2014.02.002.

- Gough, R.V., Chevrier, V. and Tolbert, M.A. *Formation of liquid water at low temperatures via the deliquescence of calcium chloride: implications for Antarctica and Mars.* Planetary and Space Science, 131:79-87, 2016, doi: 10.1016/j.pss.2016.07.006.

- Grant, W.D. *Life at low water activity.* Philosophical Transactions of the Royal Society B: Biological Sciences, 359:1249–1267, 2004, doi: 10.1098/rstb.2004.1502.

- Harri, A.-M., Genzer, M., Kemppinen, O., Gomez-Elvira, J., Haberle, R., Polkko, J., Savijärvi, H., Rennó, N., Rodriguez-Manfredi, J.A., Schmidt, W., Richardson, M., Siili, T., Paton, M., De La Torre-Juarez, M., Mäkinen, T., Newman, C., Rafkin, S., Mischna, M., Merikallio, S., Haukka, H., Martin-Torres, J., Komu, M., Zorzano, M.-P., Peinado, V., Vazquez, L. and Urqui, R. *Mars science laboratory relative humidity observations: initial results.* Journal of Geophysical Research: Planets, 119:2132–2147, 2014, doi:10.1002/2013JE004514.

- Hecht, M.H., Kounaves, S.P., Quinn, R.C., West, S.J., Young, S.M.M., Ming, D.W., Catling, D.C., Clark, B.C., Boynton, W.V., Homan, J., DeFlores, L.P., Gospodinova, K., Kapit, J. és Smith, P.H. *Detection of perchlorate and the soluble chemistry of martian soil at the Phoenix lander site.* Science, 325:64 67, 2009, doi: 10.1126/science.1172466.

- Heinz, J., Rambags, V. and Schulze-Makuch, D. Physicochemical parameters limiting growth of Debaryomyces hansenii in solutions of hygroscopic compounds and their effects on the habitability of Martian brines. Life, 11:1194, 2021, doi: 10.3390/life11111194.
- Hieta, M., Jaakonaho, I., Polkko, J., Savijärvi, H., Genzer, M., Harri, A.-M., Lorek, A., Garland, S., de Vera, J.-P., Martínez, G., Fischer, E., Sebastián Martínez, E., Rodríguez-Manfredi, J.A., Tamppari, L., de la Torre Juárez, M., McConnochie, T. REMS-H revisited: updated calibration and results of the humidity sensor of the MSL curiosity. *Space Sci. Rev.*, 221:58, 2025, doi:10.1007/s11214-025-01187-1.
- Hurowitz, J.A., et al. *Redox driven mineral and organic associations in Jezero crater, Mars. Nature*, 645: 333–345, 2025, doi: 10.1038/s41586-025-09413-0.
- Jin, Y. and Sengupta, A. Microbial habitability in porous media: pore-size constraints, transport, and colonization in micron-scale environments. Frontiers in Space Technologies, 5:1360301, 2024, doi: 10.3389/frspt.2024.1360301.
- Kuti, A. and Kereszturi, A. *Daily temperature fluctuation on Mars at aphelion.* Workshop on Planetary Atmospheres, 2007.
- Kuziakina, M., Gura, D. and Zverok, D. *GIS analysis of promising landing sites for manned flight to Mars.* E3S Web Conf. , 138:02004, 2019, doi: 10.1051/e3sconf/201913802004.
- List, Robert J. Smithsonian meteorological tables., 1951.
- de los Ríos, A., Valea, S., Ascaso, C., Davila, A.F., Kastovsky, J., McKay, C.P., Gómez-Silva, B., Wierzchos, J., 2010. Comparative analysis of the microbial communities inhabiting halite evaporites of the Atacama Desert. Int. Microbiol. 13, 79–89. doi:10.2436/20.1501.01.113.
- Marschall, M., Dulai, S. and Kereszturi, Á. *Migrating and UV screening subsurface zone on Mars as target for the analysis of photosynthetic life and astrobiology.* Planet. Space Sci., 72:146–153, 2012, doi: 10.1016/j.pss.2012.06.011.
- Martínez, G.M. and Renno, N.O. *Water and brines on Mars: current evidence and implications for MSL.* sSpace Sci. Rev, 175:29–51, 2013, doi: 10.1007/s11214-012-9956-3.
- Martín-Torres, F.J., Zorzano, M.P., Valentín-Serrano, P., Harri, A.M., Genzer, M., Kemppinen, O., Rivera-Valentin, E.G., Jun, I., Wray, J., Madsen, M.B., Goetz, W., McEwen, A.S., Hardgrove, C., Renno, N., Chevrier, V.F., Mischna, M., Navarro- González, R., Martínez-Frías, J., Conrad, P., McConnochie, T., Cockell, C., Berger, G., Vasavada, A.R., Sumner, D. and Vaniman, D. *Transient liquid water and water activity at Gale crater on Mars.* Nat. Geosci., 8:357-361, 2015, doi: 10.1038/ngeo2412.
- Maus, D., Heinz, J., Schirmack, J., Airo, A., Kounaves, S.P., Wagner, D. and Schulze-Makuch, D. *Methanogenic archaea can produce methane in deliquescence-driven Mars analog environments.* Sci. Rep., 10:6, 2020, doi: 10.1038/s41598-019-56267-4.

- McKay, C.P., Friedmann, E.I., Gómez-Silva, B., Cáceres-Villanueva, L., Andersen, D.T. and Landheim, R. *Temperature and moisture conditions for life in the extreme arid region of the Atacama Desert: four years of observations including the El Niño of 1997–1998.* Astrobiology, 3(2):393–406, 2003, doi: 10.1089/153110703769016460

- Michalski J.R., Glotch T.D., Rogers A.D., Niles P.B., Cuadros J., Ashley J.W. and Johnson S.S. (2019) *The geology and astrobiology of McLaughlin crater, Mars: an ancient lacustrine basin containing turbidites, mudstones, and serpentinites.* Journal of Geophysical Research: Planets 124, 910–940.

- Mickol, R.L., Page, J.L. and Schuerger, A.C. *Magnesium sulfate salt solutions and ices fail to protect Serratia liquefaciens from the biocidal effects of UV irradiation under Martian conditions.* Astrobiology, 17(5):387–400, 2017, doi: 10.1089/ast.2015.1448.

- Navarro, T., Madeleine, J.-B., Forget, F., Spiga, A., Millour, E., Montmessin, F. and Määttänen, A. *Global climate modeling of the Martian water cycle with improved microphysics and radiatively active water ice clouds.* Journal of Geophysical Research: Planets, 119(7):1479–1495, 2014, doi: 10.1002/2013JE004550.

- Nguyen, T., Krause, A.-C., Holmboe, M. és Yesilbas, M. *Briny water formation and retention in Martian subsurface: Insights from nontronite clay mineral and hygroscopic salts.* 55th Lunar and Planetary Science Conference, 2024.

- Nuding, D.L., Rivera-Valentín, E.G., Davis, R.D., Gough, R.V., Chevrier, V.F., Tolbert, M.A. *sDeliquescence and efflorescence of calcium perchlorate: an investigation of stable aqueous solutions relevant to Mars.* Icarus, 2014, doi: 10.1016/j.icarus.2014.08.036.

- Olsson-Francis, K. ; Cockell, C.S. 2009. *Use of Cyanobacteria for in-situ resource use in planetary exploration. J. Int. Astrobiol. Soc. 40, 557–558*

- Osano, A., Davila, A.F. *Analysis of photosynthetic activity of cyanobacteria inhabiting halite evaporites of the Atacama Desert, Chile.* Lunar and Planetary Science Conference, 2014.

- Osterloo, M.M., Hamilton, V.E., Bandfield, J.L., Glotch, T.D., Baldridge, A.M., Christensen, P.R., Tornabene, L.L. and Anderson, F.S. *Chloride-bearing materials in the southern highlands of Mars.* Science, 319(5870):1651–1654, 2008, doi: 10.1126/science.1150690.

- Osterloo, M.M., Anderson, F.S., Hamilton, V.E. és Hynek, B.M. *Geologic context of proposed chloride-bearing materials on Mars.* J. Geophys. Res., 115, 2010. doi: 10.1029/2010JE003613

- Pál, B. and Kereszturi, Á. *Annual and daily ideal periods for deliquescence at the landing site of InSight based on GCM model calculations*. Icarus, 340:113639, 2020, doi: 10.1016/j.icarus.2020.113639.

- Pál, B.D. and Kereszturi, Á. *Deliquescence probability maps of Mars and key limiting factors using GCM model calculations.* Icarus, 376:114856, 2022, doi: 10.1016/j.icarus.2021.114856.
- Pál B. and Kereszturi Á. *Possibility of microscopic liquid water formation at landing sites on Mars and their observational potential.* Icarus, 282:84-92, 2017, doi: 10.1016/j.icarus.2016.09.006.
- Pál, B., Kereszturi, Á., Forget, F., Smith, M.D., 2019. Global seasonal variation of near-surface relative humidity levels on present-day Mars. Icarus 333, 481–495. doi:10.1016/j.icarus.2019.07.007.
- Pestova, O.N., Myund, L.A., Khripun, M.K. and Prigaro, A.V. *Polythermal study of the systems $M(ClO_4)_2$–$H_2O$ ($M^{2+}$ = $Mg^{2+}$, $Ca^{2+}$, $Sr^{2+}$, $Ba^{2+}$).* Russ. J. Appl. Chem., 78:409–413, 2005, doi: 10.1007/s11167-005-0306-z.
- Rivera-Valentín, E.G., Gough, R.V., Chevrier, V.F., Primm, K.M., Martínez, G.M. and Tolbert, M.Constraining the potential liquid water environment at Gale crater, Mars., J. Geophys. Res. Planets, 123(5):1156–1167, 2018, doi: 10.1002/2018JE005558.
- Rivera-Valentín, E.G., Chevrier, V.F., Soto, A. et al. *Distribution and habitability of (meta)stable brines on present-day Mars.* Nat Astron 4, 756–761 (2020). https://doi.org/10.1038/s41550-020-1080-9
- Schofield, P.F., Knight, K.S. and Stretton, I.C. *Temperature evolution of the crystal structure of gypsum ($CaSO_4 \cdot 2H_2O$) from 4.2 K to 320 K.* Phys. Chem. Miner., 23:267–275, 1996, doi:10.1007/BF00203249.
- Schuerger, A.C. *sUltraviolet irradiation on the surface of Mars: implications for EVA activities during future human missions.* Planetary Protection Knowledge Gaps for Human Extraterrestrial Missions, 2015.
- Schuerger, A.C. and Nicholson, W.L. *Twenty species of hypobarophilic bacteria recovered from diverse soils exhibit growth under simulated Martian conditions at 0.7 kPa*. *Astrobiology*, 16(10): 786–806, 2016, doi: 10.1089/ast.2016.1587.
- Schulze-Makuch, D. *We may be looking for Martian life in the wrong place.* Nat. Astron., Comment, 2024, doi: 10.1038/s41550-024-02381-x.
- Skarbek, R.M., Savage, H.M., Kelemen, P.B. and Yancopoulos, D. *Competition between crystallization-induced expansion and creep compaction during gypsum formation, and implications for serpentinization.,* J. Geophys. Res. Solid Earth, 123:5372–5393, 2018, doi: 10.1029/2017JB015369.
- Slank, R.A., Rivera-Valentín, E.G., Chevrier, V.F., and Davila, A.F. *Salt deliquescence/efflorescence cycles in the Atacama Desert.* 55th Lunar and Planetary Science Conference (LPSC), 2024.

- Stevens, A.H. and Cockell, C.S., 2023, *The water activity of Mars-relevant multicomponent brines: the changing influence of perchlorate on habitability over time. Planet. Sci. J. 4, 6. doi:10.3847/PSJ/acaa35.*
- The Engineering ToolBox. *Liquids – Volumetric Expansion Coefficients.* The Engineering ToolBox [online], 2009.
- Titov, D.V. *Water vapour in the atmosphere of Mars.* Adv. Space Res., 29(2):183–191, 2002, doi: 10.1016/S0273-1177(01)00568-3.
- Wallace, D.C. *Thermodynamics of Crystals.,* New York: Wiley, 1972.
- Wang, A., Freeman, J.J., Ming Chou, I., and Jolliff, B.L. *Stability of $Mg^-$ sulfates at -10 $^oC$ and the rates of dehydration/rehydration processes under conditions relevant to Mars. J. Geophys. Res.*, 116, 2011, doi: 10.1029/2011JE003818..
- Wierzchos, J., Ascaso, C. and McKay, C.P. *Endolithic cyanobacteria in halite rocks from the hyperarid core of the Atacama Desert.* Astrobiology, 6(3):415–422, 2006.
- Zorzano, M.P., Mateo-Martí, E., Prieto-Ballesteros, O., Osuna, S. and Renno, N. *Stability of liquid saline water on present day Mars.* Geophys. Res. Lett., 36, 2009, doi:10.1029/2009GL040315.